%% file: mainOneColumn_Ver10.tex
\documentclass[12pt, draftclsnofoot, onecolumn]{IEEEtran}

\usepackage{cite}
\usepackage{amsmath,amssymb,amsfonts}

\usepackage{textcomp}
\usepackage{xcolor}
\usepackage{multirow}
\usepackage{makecell}
\usepackage{algorithm,algcompatible}
\usepackage{subfigure}
\usepackage{graphicx}
\usepackage{array}
\newcolumntype{P}[1]{>{\centering\arraybackslash}p{#1}}

\usepackage{booktabs}
\newtheorem{lemma}{Lemma}
\newtheorem{theorem}{Theorem}
\newtheorem{corollary}{Corollary}

\def\BibTeX{{\rm B\kern-.05em{\sc i\kern-.025em b}\kern-.08em
    T\kern-.1667em\lower.7ex\hbox{E}\kern-.125emX}}

\input{macros}

\DeclareMathOperator*{\argmax}{arg\,max}

\begin{document}


\title{Blind Massive MIMO for Dense IoT Networks}
\author{Jeongjae~Lee,~\IEEEmembership{Student Member,~IEEE}        and~Songnam~Hong,~\IEEEmembership{Member,~IEEE}
\thanks{J. Lee and S. Hong are with the Department of Electronic Engineering, Hanyang University, Seoul, Korea (e-mail: \{lyjcje7466, snhong\}@hanyang.ac.kr).}

\thanks{This work was supported in part by the Technology Innovation Program (1415178807, Development of Industrial Intelligent Technology for Manufacturing, Process, and Logistics) funded By the Ministry of Trade, Industry \& Energy(MOTIE, Korea) and in part by the Institute of Information \& communications Technology Planning \& Evaluation (IITP) under the artificial intelligence semiconductor support program to nurture the best talents (IITP-(2024)-RS-2023-00253914) grant funded by the Korea government(MSIT).}
}



\maketitle

\begin{abstract}

In this paper, we investigate the downlink communication challenges in heavy-load Internet-of-Things (IoT) networks supported by frequency-division-duplexing (FDD)  millimeter-wave (mmWave) massive multiple-input multiple-output (MIMO) systems. The excessive overhead required for obtaining channel state information at the transmitter (CSIT) is essential to achieve high spectral efficiency through conventional massive MIMO techniques; however, it hinders the deployment of ultra-reliable low-latency communications (URLLC) and leads to significant energy expenditure, particularly in dense IoT networks. To address this challenge, we propose an innovative CSIT-Free MIMO precoding method, referred to as CIRculant information Classification via Linear Estimation (CIRCLE). Our major contribution is the design of a CSIT-independent (or deterministic) precoding, which is constructed by leveraging the circulant permutation of the discrete Fourier transform (DFT) matrix. This design enables interference-free signal combining at the IoT devices. Through theoretical analysis and simulations, we verify the effectiveness of the proposed CIRCLE method.

\end{abstract}

\section{Introduction}\label{sec:intro}
Multi-user communication in dense wireless networks has received considerable attention due to its efficient resource utilization. The emergence of heavy payload Internet-of-Things (IoT) applications in dense networks, such as virtual reality, Internet-of-Drones (IoD), and video messaging in automotive contexts, has heightened the demand for exceptionally high-quality service among IoT devices \cite{wu2023, Kim2019,Yazdine2021}. This demand encompasses not only ultra-reliable and low-latency communications (URLLC) \cite{Osseiran2014,Durisi2016}, which are essential for real-time applications, but also the need for high spectral efficiency. To meet these stringent requirements, extensive research has been conducted on massive multiple-input multiple-output (MIMO) systems \cite{marzetta2010noncooperative, Ngo2013}. These systems play a crucial role in achieving enhanced spectral efficiency. Moreover, the integration of higher frequency bands, such as millimeter-wave (mmWave) and terahertz (THz), is expected to yield substantial bandwidth resources, thereby improving overall network performance \cite{wang2018millimeter, Elayan2020}. In dense IoT networks supported by massive MIMO, low energy operation frequently takes precedence over metrics such as latency and data rates \cite{Lopez2020}. This prioritization arises from the reliance of simple IoT devices on battery power and/or limited energy sources, as exemplified in wireless-powered communication networks. A significant portion of energy consumption is attributed to the acquisition of channel state information at the transmitter (CSIT), which is essential for enhancing spectral efficiency. The accuracy of CSIT is directly related to the training overhead, including pilot or feedback transmission. Consequently, a trade-off exists among energy expenditure, latency, and spectral efficiency. Nevertheless, the challenge of minimizing latency while ensuring low energy consumption and maintaining high spectral efficiency remains critical for the effective deployment of massive MIMO in dense IoT networks. This is the motivation of our work.

In massive MIMO systems utilizing high-frequency bands, frequency-division-duplexing (FDD) emerges as a viable option for deploying 
URLLC with low energy consumption, as it leverages abundant bandwidth to mitigate interference between simultaneous uplink and downlink communications. In downlink communications, FDD massive MIMO can achieve high spectral efficiency through well-established precoding methods utilizing CSIT (in short, CSIT-Based methods) \cite{Jindal2006,Love2005,Maddah2012,Yang2013}. However, acquiring CSIT is complex due to the non-reciprocal nature of uplink and downlink channels, which necessitates CSIT feedback. This feedback incurs substantial overhead, thereby impeding the low-latency operation of FDD massive MIMO systems. Consequently, this requirement leads to a significant increase in energy consumption for IoT devices.

Numerous studies have been conducted to mitigate the overhead associated with acquiring CSIT in FDD massive MIMO systems \cite{Huang2009,Lee2011,Gao2016,Han2017,Barzegar2019,Han2024,Kim2024}. Despite these efforts, channel state information (CSI) feedback or uplink pilot transmission remains necessary, resulting in considerable energy consumption at the IoT devices. In response to this challenge, a downlink transmission strategy that operates without acquiring CSIT may be considered, relying instead on channel state information at the receiver (CSIR) to detect the desired symbols. This approach can eliminate the need for CSI feedback and uplink pilot transmission, thus addressing the stringent ultra-low latency requirements inherent to URLLC with low energy consumption. To the best of our knowledge, this strategy has garnered limited attention in the literature, primarily due to perceived performance limitations \cite{Zhu2009}. Nevertheless, there exists potential for realizing downlink transmission without CSIT by leveraging  space-time block coding (STC) \cite{Alamouti1998,Tarokh1999,Arti2020}. Notably, this method is restricted to single-device downlink transmission, which constrains its scalability and applicability in scenarios involving multiple devices within dense IoT networks. In this paper, we advance the concept of STC, demonstrating that interference-free combining among multiple IoT devices is achievable through the proposed CSIT-Free precoding method based on time-sequential downlink transmissions.




A limited CSI feedback approach was introduced in \cite{Huang2009,Lee2011}, demonstrating that quantized CSI feedback can significantly minimize overhead. The compressed sensing (CS) framework presented in \cite{Gao2016,Han2017} diminishes the number of feedback parameters by leveraging the inherent sparsity of MIMO channels. Additionally, a promising method introduced in \cite{Barzegar2019,Han2024,Kim2024} utilizes uplink pilots for CSIT reconstruction, achieving notable performance and low latency. Specifically, in \cite{Han2024}, the central concept involves exploiting frequency-independent channel parameters estimated from uplink pilots, such as the angle of departure (AoD) and the physical distance between the transmitter and receiver, to reconstruct the downlink channel. 
Recently, \cite{Kim2024} improved the performance of downlink channel reconstruction by incorporating limited CSI feedback. However, this method results in several impractical issues.  Firstly, substantial overhead and complexity in uplink channel estimation are required, particularly as the number of antennas and devices increase. In the aforementioned works, uplink channels are typically assumed to be perfectly estimated, which may not be feasible in practical scenarios.
Additionally, the sequence of uplink pilot transmissions, uplink channel estimation, downlink channel reconstruction, and derivation of precoding solutions leads to significant latency and computational complexity. These challenges are particularly pronounced when deploying massive MIMO systems in dense IoT networks. More critically, transmitting data symbols to IoT devices within a short coherence time becomes infeasible, presenting a significant obstacle to the implementation of URLLC. 
Furthermore, existing works often rely on suboptimal precoding methods, such as zero-forcing (ZF) or minimum mean square error (MMSE), utilizing the estimated DL channels. These suboptimal approaches may underperform compared to an optimal performance, particularly when signal-to-noise (SNR) or channel estimation accuracy diminishes. This ultimately leads to the overutilization of resources in massive MIMO systems \cite{Zhao2015}.

The most pertinent work related to this paper involves the transmission of desired information symbols without reliance on CSIT. By eliminating the need for CSIT, this approach paves the way for more efficient communication strategies in various URLLC applications. In \cite{Zhu2009}, the authors derived an outer bound for the degree-of-freedom (DoF) region of a two-user MIMO interference isotropic fading channel, assuming channel state information is known only at the receivers. They demonstrated that the absence of CSIT results in a loss of DoF, underscoring the importance of CSIT in enhancing DoF performance. As another approach, STC was introduced in \cite{Alamouti1998, Tarokh1999}, utilizing CSI-Free precoding at the transmitter, constructed through time-sequential complex orthogonal codes. This method allows the receiver to effectively detect data symbols and achieve maximum ratio transmission (MRT) performance. While STC demonstrates notable performance with minimal transmitter processing \cite{Tarokh1999}, it is limited to single-user communication aimed at diversity gain. To enable multi-user communication,  \cite{Arti2020} proposed a novel downlink data transmission method that operates without CSIT, based on STC, which customizes the interference alignment (IA) method for effective interference cancellation and desired information detection. However, achieving satisfactory spectral efficiency requires a substantial number of antennas at the receivers, which increases  system complexity. This poses significant challenges for deployment in IoT systems, which typically consist of single-antenna devices.

In this paper, we investigate the downlink data transmission in dense IoT networks characterized by heavy payloads and supported by FDD massive MIMO. Within this framework, we propose an innovative precoding method that operates without the need for CSIT, referred to as CSIT-Free precoding. This method effectively addresses the stringent ultra-low latency requirement of URLLC and ensures the low energy efficiency necessary for IoT devices. Our major contributions are summarized as follows:
\begin{itemize}
    \item We consider dense IoT networks in which a base station (BS) equipped with a massive number of antennas, $N$, serves $K=N-2$ IoT devices. Notably, there is no feedback channel or uplink pilots available for transmitting CSI information to the BS, in order to meet the stringent latency requirements of URLLC and adhere to low-power constraints for IoT devices. To address this challenge, we propose a CSIT-Free MIMO precoding method, referred to as {\bf CIR}culant information {\bf C}lassification via {\bf L}inear {\bf E}stimation ({\bf CIRCLE}).

    \item The proposed CIRCLE method represents an innovative deterministic (or CSIT-Free) precoding approach, constructed by utilizing the circulant permutation of the DFT matrix. As proved in Lemma 1, the resulting precoding matrices can be regarded as orthogonal bases when integrated with the proposed combining method at the receivers. Leveraging this property, each IoT device is capable of perfectly canceling interference signals (refer to Lemma 2), thereby achieving maximum spectral efficiency, provided that the channel is accurately estimated at the device.


    \item We introduce an efficient channel estimation algorithm specifically designed for our CIRCLE method. This algorithm requires only two pilot symbols per time slot, which are transmitted concurrently with $N-2$ data symbols. By leveraging the predominance of the line-of-sight (LoS) channel in mmWave communications, each device can independently acquire its channel by solving the maximization problem based on the proposed objective function (for detailed information, refer to Lemma 3 and Corollary 1). Furthermore, by extending our channel estimation algorithm to wideband orthogonal frequency division multiplexing (OFDM) systems, we can significantly enhance channel estimation accuracy through joint optimization across the subcarriers, thereby improving achievable spectral efficiency.
    

    \item Finally, we present Theorem 1 as the main result of this paper, asserting that the proposed CIRCLE method, utilizing estimated CSIR, can nearly achieve maximum sum-spectral efficiency under certain conditions. Our simulations demonstrate that the CIRCLE method exhibits remarkable performance, closely approaching the performance bound, thus validating our theoretical analysis presented in Theorem 1. Furthermore, the CIRCLE method outperforms CSIT-Based benchmark methods in terms of both performance, specifically sum-spectral efficiency, and latency. Consequently, it emerges a strong candidate for a CSIT-Free MIMO precoding method suitable for dense IoT networks that require ultra-low latency and energy efficiency.

    
\end{itemize}

\begin{figure*}[t]
\centering
\includegraphics[width=1.0\linewidth]{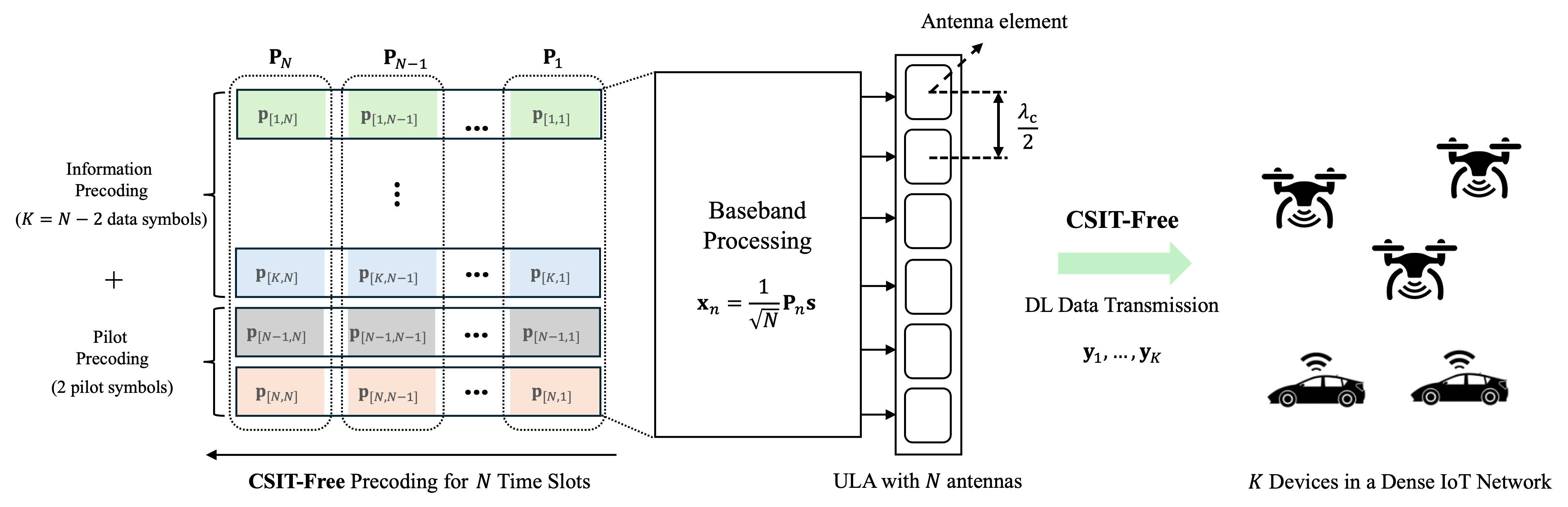}
\caption{Description of the proposed DL data transmission without CSIT in heavy payload IoT communication systems supported by FDD based mmWave massive MIMO system.}
\end{figure*}
{\em Notations.} Let $[N_1: N_2]\eqdef \{N_1,N_1+1,...,N_2\}$ for any integer $N_1, N_2$ with $N_2>N_1$. For $N_1=1$, this notation simplifies to $[N_2]$. We use $\xv$ and $\Am$ to denote a column vector and matrix, respectively. Given a $M \times N$ matrix $\Am$, let $\Am(i,:)$ and $\Am(:,j)$ denote the $i$-th row and $j$-th column of $\Am$, respectively. Given $m < M$ and $n < N$, let $\Am([m],:)$ and $\Am(:,[n])$ denote the submatrices by taking the first $m$ rows and $n$ columns of $\Am$, respectively. Also, $\Am^{\herm}$ and $\|\Am\|_F$ represent the Hermitian transpose and the Frobenius norm of $\Am$, respectively. Given a vector $\vv$, ${\vv}^{*}$ denotes the complex conjugate vector of $\vv$ and ${\rm diag}(\vv)$ denotes a diagonal matrix whose $\ell$-th diagonal element corresponds to the $\ell$-th element of $\vv$.

\section{Preliminaries}

We consider heavy payload Internet-of-Things (IoT) communication systems \cite{Kim2019} supported by frequency-division-duplexing (FDD) based mmWave massive multiple-input multiple-output (MIMO) systems. In this context, the base station (BS) employs a uniform linear array (ULA) consisting of $N$ antennas. Assuming a dense IoT network in which $K$ devices are randomly and uniformly distributed across the network, the BS serves single-antenna equipped devices within a single-cell, as illustrated in Fig. 1. For downlink data transmission within the FDD massive MIMO framework, the BS operates at a central carrier frequency, denoted as $f_{\rm c}$. The wavelength of the central carrier, denoted as $\lambda_{\rm c}$, is given by $\lambda_{\rm c}=c/{f_{\rm c}}$,  where $c$ represents the speed of light. The antenna spacing of the ULA at the BS is configured to be half of the central wavelength, specifically, $\lambda_{\rm c}/2$. In this paper, we focus on downlink data transmission without the knowledge of channel state information at the transmitter (CSIT). To this end, we propose innovative precoding and combining methods to efficiently support a large number of densely situated devices in URLLC IoT networks. The subsequent subsections will articulate the signal model employed, along with the proposed downlink precoding structures. 

\subsection{Signal Model}

The downlink channel response from the BS to the $k$-th device is defined as follows:
\begin{equation}
    \hv_k = \alpha_k\av\left(\theta_k\right) + \sum_{\ell=1}^{L_k}\beta_{[k,\ell]}\av\left(\vartheta_{[k,\ell]}\right)\in\CC^{N\times 1},\label{eq:channel}
\end{equation} for $k \in [K]$. Regarding the line-of-sight (LoS) path, $\alpha_k\in\CC$ and $\theta_k\in[0,2\pi)$ denote the complex channel gain and the angle-of-departure (AoD) from the BS to the $k$-th device, respectively. For each $\ell\in[L_k]$, $\beta_{[k,\ell]} \in \CC$ and $\vartheta_{[k,\ell]}\in[0,2\pi)$ denote the complex channel gain and the AoD in $\ell$-th non line-of-sight (NLoS) paths, respectively. Also, $\av(\theta)$ represents the array response vector, defined as follows:
\begin{equation}
    \av(\theta) = \left[1,e^{j{\pi}\sin{\theta}},\dots,e^{j\pi(N-1)\sin{\theta}}\right]^{\intercal}.
\end{equation} Considering the downlink channel as described in \eqref{eq:channel}
the received signal model for the $k$-th user can be expressed as follows:
\begin{equation}
    y_k = \sqrt{p_{\rm t}}\hv_k^{\herm}\xv + z_k,\label{eq:received}
\end{equation} where $\xv\in\CC^{N\times 1}$ represents the transmit signal vector, which is normalized such that $\|\xv\|_2^2=1$, $p_{\rm t}$ denotes the transmit power, and $z_k\sim \Cc\Nc(0,\sigma^{2})$ indicates the additive Gaussian white noise with zero means and variance $\sigma^2$.


\subsection{The Proposed CSIT-Free Precoding Method}\label{subsec:precoding}
We consider a communication scenario in which devices, such as automotive vehicles or drones, frequently change their locations. This mobility results in a short coherence time for the downlink channels. In this context, it becomes particularly challenging for the devices to transmit CSIT feedback, as well as for the BS to reconstruct the downlink channels based on uplink pilots. This difficulty is compounded by the need to perform CSIT-Based downlink transmission within a limited coherence time.  Consequently, this conventional transmission strategy may not be adequate for ultra-reliable and low-latency communications (URLLC), which require consistent and timely data transmission. Moreover, as devices are required to transmit signals to the BS in any way, energy consumption becomes significantly elevated. In mmWave communications, especially, the use of high-frequency bands results in substantial energy expenditure at the devices, where the transmit power of IoT device is considerably constrained. These limitations of traditional methods necessitate the exploration of alternative transmission strategies that can accommodate the dynamic nature of user mobility while fulfilling the stringent requirements of URLLC and low-energy applications. Motivated by this, we propose a novel downlink precoding method that is constructed without the knowledge of CSIT, referred to as CSIT-Free precoding method, along with a combining method for each device. This method facilitates an {\em interference-free} transmission, thereby enabling each device to reliably detect the desired information. As illustrated in Fig. 1, the BS is capable of transmitting new information to the $K$ devices during the $N$ time slots, without the knowledge of CSIT. During these $N$ time slots, the concatenation of the received signals at the $k$-th device is represented as follows:
\begin{equation}
    \yv_k = \begin{bmatrix}
        y_{[1,k]} &\cdots & y_{[N,k]}
    \end{bmatrix}^{\intercal}\in\CC^{N\times 1},\label{eq:concrsignal}
\end{equation} where the received signal at the $n$-th time slot is given by
\begin{equation}
    y_{[n,k]} = \sqrt{p_{\rm t}}\hv_k^{\herm}\xv_n + z_{[n,k]},
\end{equation} and where $\xv_n\in\CC^{N\times 1}$ denotes the downlink transmission vector at the $n$-th time slot and $z_{[n,k]}\sim\Cc\Nc(0,\sigma^{2})$. Here, the DL transmission vector at the $n$-th time slot is constructed as follows:
\begin{equation}
    \xv_n =  \frac{1}{\sqrt{N}}\Pm_n\sv,\label{eq:transmissionvector}
\end{equation} where $\Pm_{n}\in\CC^{N\times N}$ is the CSIT-Free (or deterministic) precoding matrix and $\sv\in\CC^{K\times 1}$ denotes the symbol vector satisfying $\EE[\sv\sv^{\herm}]=\Id_N$. Note that the symbol vector $\sv$ comprises the $K$ information symbols associated with the devices in $\sv([K])$ and $N-K$ pilot symbols used to estimate channels in $\sv([K+1:N])$.
Let $\pv_{[n,k]}$ denote the $k$-th column of the matrix $\Pm_{n}$. We construct the linear combiner at the $k$-th device as follows:
\begin{align}
    {\Fm}_k = \begin{bmatrix}
        \pv_{[1,k]} &\cdots & \pv_{[N,k]}
    \end{bmatrix} = \Um_{[N,k]}\in\CC^{N\times N},\label{eq:precoding}
\end{align} where $\Um_{[N,n]}$ is defined in Lemma 1. Notably, the proposed precoding matrices $\{\Pm_n: n \in [N]\}$ and linear combining matrices $\{\Fm_{k}: k \in [N]\}$ are constructed without the knowledge of CSIT and posses unique structures derived from permuted DFT matrices. This design is specifically intended to eliminate inter-device interferences. 

It is noteworthy that only two pilot symbols are sufficient to estimate the downlink channels at the receivers, although our framework is more general. Consequently, the proposed method can effectively serve up to $K=N-2$ devices (or users). In the context of massive MIMO systems, this efficiency renders the pilot overhead negligible, as the ratio $(N-2)/N$ approaches 1. Therefore, our method represents a promising candidate for effectively serving numerous devices in IoT networks supported by FDD massive MIMO.

\vspace{0.1cm}
\begin{lemma}
    Let an $N\times N$ circulant matrix $\Cm_N$ be given as follows:
    \begin{equation}
        \Cm_N \eqdef \begin{bmatrix}
            1 & N &\cdots & 2  \\
            2 & 1 & \ddots & \vdots  \\
            \vdots & \vdots & \ddots & N \\
            N & N-1 & \cdots & 1 
        \end{bmatrix}.\label{eq:circulant}
    \end{equation} Given an $N\times N$ normalized discrete Fourier transform (DFT) matrix $\Um_N$, we define the $N$ permutation matrices of $\Um_N$, where the $k$-th permutation matrix is defined as follows:
    \begin{equation}
        \Um_{[N,k]} \eqdef \begin{bmatrix}
            \Um_{N}(:,\Cm_N(1,k)) &\cdots& \Um_{N}(:,\Cm_N(N,k))
        \end{bmatrix},\label{eq:permutation}
    \end{equation} for $k \in [N]$. Then, it holds that
    \begin{equation}\label{eq:diffuser}
        \Um_{[N,k]}\Um_{[N,k']}^{\herm} = \begin{cases}
            \Id_N, & k=k'\\
            \mbox{diag}(\uv_{[k,k']}), & k \neq k'
        \end{cases}
    \end{equation} where $\uv_{[k,k']}$ is a vector such that ${\bf 1}_N^{\herm}\uv_{[k,k']} = 0$.
\end{lemma}
\begin{IEEEproof}
    The proof is provided in Appendix A.
\end{IEEEproof} 

In the subsequent section, we will delineate an efficient method for detecting symbols solely from the received signal vector presented in \eqref{eq:concrsignal}. This approach leverages the unique structures inherent in the precoding and combining matrices.

\section{Method}

As the central focus of this paper, we present how to accurately detect the desired symbols from the received signal vector $\{\yv_k:k\in[K]\}$ processed by the proposed CSIT-Free downlink precoding. We will first elucidate the detection procedures of the proposed method when the channel state information at the receiver (CSIR) is fully available. Subsequently, in Section~\ref{subsec:formulation}, we will formulate the channel estimation problem to facilitate the proposed method in practice. In Section~\ref{subsec:channel1} and~\ref{subsec:wideband}, the proposed channel estimation algorithms will be detailed using the received signal, followed by a theoretical analysis of our channel estimation approach.

\subsection{Circulant Information Classification via Linear Estimation}

Focusing on the $k$-th device, we elucidate the proposed downlink transmission method, referred to as  {\bf CIR}culant information {\bf C}lassification via {\bf L}inear {\bf E}stimation ({\bf CIRCLE}), along with its theoretical analysis. The same procedures will subsequently be applied to the other devices thanks to the circulant symmetry of the proposed CIRCLE method. Assuming that $\hv_k$ is perfectly estimated (i.e., full-CSIR), the $k$-th device can leverage the linear combiner $\Fm_k$ to obtain the following: 
\begin{align}
    d(\hv_k,{\Fm}_k,\yv_k) \eqdef \tilde{\hv}_k^{\transp}{\Fm}_k^{*}\yv_k\in\CC,\label{eq:linearreceiver}
\end{align} where
\begin{equation} 
    \tilde{\hv}_k = {\bf 1}_N\oslash(\hv_k^{*}),\label{eq:elementwisedividing}
\end{equation} and $\oslash$ denotes the component-wise division. We will show that the above linear combiner can completely cancel the interference signals harnessing the special property of ${\Fm}_k$, as provided in Lemma 1. To specify the interference canceling process, we divide $d(\hv_k,{\Fm}_k,\yv_k)$ into three distinct terms as follows:
\begin{align}
    &d(\hv_k,{\Fm}_k,\yv_k)= \sv(k){\sqrt{\frac{p_{\rm t}}{N}}}\tilde{\hv}_k^{\transp}{\sum_{n=1}^{N}\pv_{[n,k]}^{*}\hv_{k}^{\herm}\pv_{[n,k]}}\nonumber\\
    &\quad + \sv(k'){ \sqrt{\frac{p_{\rm t}}{N}}}\tilde{\hv}_k^{\transp}\sum_{k'\neq k}^{N}{\sum_{n=1}^{N}\pv_{[n,k]}^{*}\hv_{k}^{\herm}\pv_{[n,k']}}+ \tilde{\hv}_k^{\transp}{\Fm}_k^{*}\zv_k\nonumber\\
    &=\sv(k)\sqrt{\frac{p_{\rm t}}{N}}g(\hv_k,{\Fm}_k,\yv_k) + \sqrt{\frac{p_{\rm t}}{N}}\sum_{k'\neq k}^{N}\sv(k')v(\hv_k,{\Fm}_{k'},\yv_k)\nonumber\\
    &\quad + \tilde{\hv}_k^{\transp}{\Fm}_k^{*}\zv_k,\label{eq:decoded}
\end{align} where $g(\hv_k,{\Fm}_k,\yv_k)$ and $v(\hv_k,{\Fm}_{k'},\yv_k)$ are defined in \eqref{eq:cgdsignal} and \eqref{eq:cgisignal}, respectively, and the concatenation of the additive noises $\{z_{[i,k]}:i\in[N]\}$ is defined as follows:
\begin{equation}
\zv_k=\left[z_{[1,k]},z_{[2,k]},\dots,z_{[N,k]}\right]^{\intercal}\in\CC^{N\times 1},
\end{equation} with $\EE[\zv_k\zv_k^{\herm}]=\sigma^2\Id_N$. In the combined signal in \eqref{eq:decoded}, the first term is associated with the desired symbol, i.e., $\sv(k)$, while the second term encompasses interference signals. Notably, our key findings are provided in Lemma~\ref{lem:linear_combiner} below:
\begin{lemma}\label{lem:linear_combiner} Leveraging the proposed CSIT-Free linear precoding and the linear receiver in \eqref{eq:linearreceiver}, the desired and interference signals are described as follows:
\begin{align}
     g(\hv_k,{\Fm}_k,\yv_k) &= \tilde{\hv}_k^{\transp}\sum_{n=1}^{N}\pv_{[n,k]}^{*}\hv_{k}^{\herm}\pv_{[n,k]} = N\label{eq:cgdsignal}\\ 
     v(\hv_k,{\Fm}_{k'},\yv_k)&= \tilde{\hv}_k^{\transp}\sum_{n=1}^{N}\pv_{[n,k]}^{*}\hv_{k}^{\herm}\pv_{[n,k']}= 0.\label{eq:cgisignal}
\end{align} 
\end{lemma}
\begin{IEEEproof} We first prove that the combining gain of the desired signal is obtained as $N$ as follows:
\begin{align}
    g(\hv_k,{\Fm}_k,\yv_k) = \tilde{\hv}_{k}^{\transp}\Fm_{k}^{*}\Fm_{k}^{\transp}\hv_{k}^{*}&\stackrel{(a)}{=}\tilde{\hv}_{k}^{\transp}\left(\Fm_{k}\Fm_{k}^{\herm}\right)^{*}\hv_{k}^{*}\nonumber\\
    &\stackrel{(b)}{=}\tilde{\hv}_{k}^{\transp}\hv_{k}^{*}\stackrel{(c)}{=} N,\label{eq:signal}
\end{align} where (a) is due to the fact that $(\Am\Bm)^{*}=\Am^{*}\Bm^{*}$ for some complex matrices $\Am$ and $\Bm$, (b) is from Lemma 1, and (c) follows from \eqref{eq:elementwisedividing}. We next show that the interference signals are completely canceled as follows:
\begin{align}
    v(\hv_k,{\Fm}_k',\yv_k) = \tilde{\hv}_{k}^{\transp}\Fm_{k}^{*}\Fm_{k'}^{\transp}\hv_{k}^{*}&=\tilde{\hv}_{k}^{\transp}(\Fm_{k}\Fm_{k'}^{\herm})^{*}\hv_{k}^{*}\nonumber\\
    &\stackrel{(a)}{=}\tilde{\hv}_{k}^{\transp}(\mbox{diag}(\uv_{[k,k']}))^{*}\hv_{k}^{*}\nonumber\\
    &=\tilde{\hv}_{k}^{\transp}\mbox{diag}(\hv_{k}^{*})\uv_{[k,k']}^{*}\nonumber\\
    &\stackrel{(b)}= 0,\label{eq:isignal}
\end{align} where (a) is from Lemma 1, and (b) is due to the fact that $\tilde{\hv}_{k}^{\transp}\mbox{diag}(\hv_{k}^{*}) = {\bf 1}_N$ from the equation in \eqref{eq:elementwisedividing} and ${\bf 1}_{N}^{\herm}\uv_{[k,k']} = ({\bf 1}_{N}^{\transp}\uv_{[k,k']}^{*})^{*}=0$ from Lemma 1. This completes the proof.
\end{IEEEproof}
\vspace{0.1cm}
Lemma~\ref{lem:linear_combiner} shows that our linear combiner completely cancels the interference signals while achieving the full combining gain for the desired signal. Consequently, the combined signal in \eqref{eq:decoded} is expressed as follows:
\begin{align}
    d(\hv_k,{\Fm}_k,\yv_k) = {\sqrt{p_{\rm t}N}}\sv(k)+ \tilde{\hv}_k^{\transp}{\Fm}_k^{*}\zv_k.\label{eq:cancel}
\end{align}
We finally complete the explanation of the proposed interference cancellation process.

\begin{figure}[t]
\centering
\includegraphics[width=0.5\linewidth]{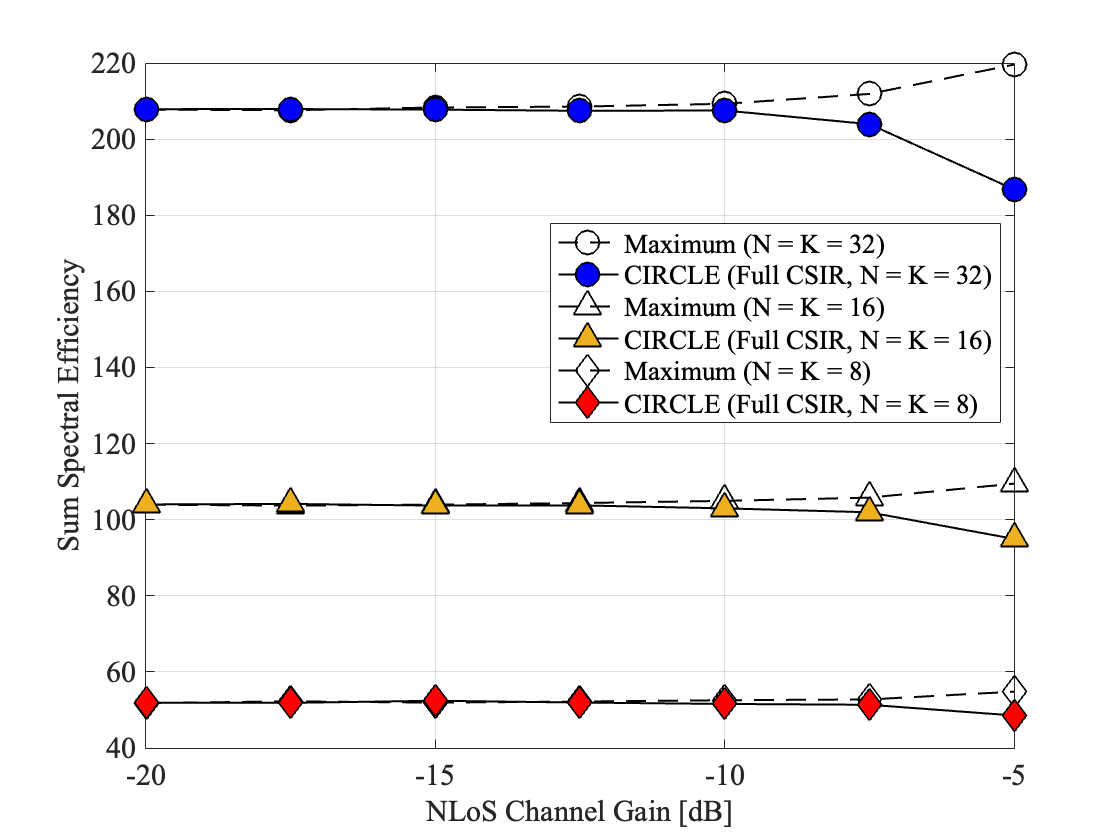}
\caption{The sum-spectral efficiency of the proposed linear method (named CIRCLE) as a function of the NLoS channel gain.}
\end{figure}

Since the interference is perfectly canceled by the proposed linear method, noticeably, the corresponding signal-to-interference-plus-noise ratio (SINR) is computed via \eqref{eq:cancel}:
\begin{align}
    {\gamma}^{\star}({\hv}_k,{\Fm}_k,\yv_k) &= \frac{\EE\left[|\sqrt{p_{\rm t}N}\sv(k)|^2\right]}{\EE[|\tilde{\hv}_k^{\transp}{\Fm}_k^{*}\zv_k|^2]}\nonumber\\
    &\stackrel{(a)}{=}\frac{p_{\rm t}N}{\sigma^2\tilde{\hv}_k^{\transp}\tilde{\hv}_k^{*}} \stackrel{(b)}{\leq} \frac{p_{\rm t}{\hv}_k^{\transp}{\hv}_k^{*}}{N\sigma^2},\label{eq:maxSINR}
\end{align} where (a) comes from the fact that $\EE[\zv_k\zv_k^{\herm}]=\sigma^2\Id_N$, (b) follows from the relation between the harmonic mean and the arithmetic mean, and the equality in (b) holds if and only if the diagonal elements of $\hv_k^{*}\hv_k^{\transp}$ are all equal. In the upper bound, remarkably, the desired signal is maximized by fully achieving the channel gain ${\hv}_k^{\transp}{\hv}_k^{*}$. It follows that the upper bound represents the maximum achievable SINR of the linear method. In mmWave communications, the LoS path is predominant due to the severe path loss of the NLoS paths. Consequently, the diagonal elements of $\hv_k^{*}\hv_k^{\transp}$ are likely to be nearly equal. Therefore, this suggests that the proposed method can nearly achieve the upper bound by perfectly canceling the interference from other devices, provided that $\hv_k$ is perfectly estimated. Based on this, the proposed method can nearly achieve the maximum sum-spectral efficiency, defined as:
\begin{align}
    \Rc^{\star} 
    &\eqdef \sum_{k=1}^{K}\log_2\left(1+\frac{p_{\rm t}{\hv}_k^{\transp}{\hv}_k^{*}}{N\sigma^2}\right).\label{eq:maxRATE}
\end{align} 

To support this argument, we evaluate the performance of the proposed CIRCLE method. Fig. 2 illustrates its sum-spectral efficiency, denoted as $\Rc^{\star}$, as a function of the channel gain of the NLoS paths, represented by $\delta^2$. For the simulations, we assume that the complex channel gain of the LoS path follows a normal distribution with zero mean and unit variance, i.e., $\alpha_k\sim\Cc\Nc(0,1)$ for $k \in [K]$. Additionally, we assume that $f_{\rm c} = 100$ GHz, $p_{\rm  t} = 0$ dB, $\sigma^2 = -10$ dB, and $L_k = 3$, where the complex channel gain of the NLoS paths is defined as $\beta_{[k,\ell]}\sim\Cc\Nc(0,\delta^2)$. We observe that the proposed CIRCLE method can achieve the maximum sum-spectral efficiency, as defined in \eqref{eq:maxRATE} (referred to as ``Maximum'' in the legend in Fig. 2), when the NLoS channel gain, denoted as $\delta^2$, decreases. This observation confirms that the CIRCLE method is indeed capable of fully eliminating interference. However, a performance gap persists between the maximum sum-spectral efficiency and the achievable sum-spectral efficiency of the CIRCLE method when the NLoS channel gain is relatively high (e.g., $\delta^2 = -10 \sim -5$ dB). This gap is attributed to the fact that the SINR in \eqref{eq:maxSINR} fails to achieve the upper bound, as the diagonal elements of $\hv_k^{*}\hv_k^{\transp}$ differ due to the effects of the NLoS paths. Notably, in mmWave communications, significant reflection losses lead to NLoS channels experiencing average attenuation that exceeds $10$ dB compared to the LoS channel \cite{Priebe2013}. This attenuation can be expressed as $\delta^2 < -10$ dB.
Consequently, we assert that under full-CSIR, the proposed CIRCLE method can nearly achieve the maximum sum-spectral efficiency in mmWave communications.

\subsection{Channel Estimation Problem}\label{subsec:formulation}
However, in practice, the aforementioned maximum sum-spectral efficiency cannot be achieved when channel estimation errors at the devices exist. In this case, complete interference cancellation becomes unattainable, resulting in significant degradation of the SINR. Thus, it is imperative for the devices to accurately estimate the channels before conducting the proposed detection procedures. To estimate $\hv_{k}$, one might consider utilizing additional downlink pilots, albeit at the cost of increased communication latency. We address this challenge by presenting an efficient method to accurately estimate the channels using only the received signal vector $\yv_{k}$. Motivated by the predominance of the LoS path in mmWave systems, the proposed channel estimation method is developed under the assumption that $\hv_{k}=\alpha_{k}\av(\theta_k)$, while neglecting the impact of the NLoS paths in \eqref{eq:channel}. The implications of NLoS paths will be thoroughly examined in detail in our simulations.

From \eqref{eq:linearreceiver} and \eqref{eq:decoded}, the combined signal using the estimated channel $\hat{\hv}_k$ is represented as follows:
\begin{align}
    &d(\hat{\hv}_k,{\Fm}_k,\yv_k) = \hat{\hv}_k^{\intercal}{\Fm}_k^{*}\yv_k=\sqrt{p_{\rm t}}\sv(k)g(\hat{\hv}_k,{\Fm}_k,\yv_k) \nonumber\\
    &\quad\quad\quad\quad+ \sqrt{p_{\rm t}}\sum_{k'\neq k}^{N}\sv(k')v(\hat{\hv}_k,{\Fm}_{k'},\yv_k)+\hat{\hv}_k^{\intercal}{\Fm}_k^{*}\zv_k.
\end{align} Accordingly, the SINR is computed as:
\begin{equation}
    \gamma(\hat{\hv}_k,{\Fm}_k,\yv_k) = \frac{\EE\left[|{P}_k|^2\right]}{\EE\left[|{I}_k|^2\right]},\label{eq:SINR}
\end{equation} where the desired signal and the interference-plus-noise terms are respectively given by
\begin{align}
    {P}_{k} &= \sqrt{p_{\rm t}}\sv(k)g(\hat{\hv}_k,{\Fm}_{k},\yv_k),\\
    {I}_{k} &= \sqrt{p_{\rm t}}\sum_{k'\neq k}^{N}\sv(k')v(\hat{\hv}_k,{\Fm}_{k'},\yv_k)+\hat{\hv}_k^{\intercal}{\Fm}_k^{*}\zv_k.
\end{align} From \eqref{eq:cancel}, it is evident that the SINR in \eqref{eq:SINR} approaches the maximum SINR, denoted as $\gamma^{\star}({\hv}_k,{\Fm}_k,\yv_k)$, as the estimated channel $\hat{\hv}_k$ closely aligns with the true channel $\hv_k$. This alignment effectively removes interference signals and maximizes the desired signal power. Harnessing this fact, we can estimate a channel $\hv_{k}=\alpha_k \av(\theta_k)$ by taking the SINR $\gamma(\hv=\alpha\av(\theta),\Fm_k,\yv_k)$ as an objective function. To this end, we provide a key lemma below:
\vspace{0.1cm}
\begin{lemma}
    Given the complex channel gain $\alpha_k$ of $\hv_k$, we let $\phi^{\star}$ denote the global maximum of $\gamma(\alpha_k\av(\phi),{\Fm}_k,\yv_k)$, i.e.,
    \begin{equation}
        \phi^{\star} = \argmax_{\phi \in [-\pi, \pi)} \gamma(\alpha_k\av(\phi),{\Fm}_k,\yv_k).\label{eq:lem-opt}
    \end{equation} Then, our solution ensures that
    \begin{equation}
        \av(\phi^{\star}) = \av(\theta_k).
    \end{equation}
\end{lemma}
\begin{IEEEproof}
    The proof is provided in Appendix C.
\end{IEEEproof} 
\begin{corollary} For any $\hat{\alpha} \in \CC$, we let
\begin{equation}
    \bar{\phi}^{\star} = \argmax_{\phi \in [-\phi,\phi)} \gamma(\hat{\alpha} \av(\phi),{\Fm}_k,\yv_k).\label{eq:opt1}
\end{equation} This solution also ensures that 
\begin{equation}
        \av(\bar{\phi}^{\star}) = \av(\theta_k).
\end{equation}
\end{corollary}
\begin{IEEEproof}
    The proof is derived by the proof of Lemma 3.
\end{IEEEproof}

\vspace{0.1cm}
Based on Lemma 3, we can estimate the array response vector, $\av(\theta_k)$, by solving the maximization problem specified in \eqref{eq:lem-opt}. Notably, Corollary 1 suggests that this estimation remains stable despite potential estimation errors in the complex channel gain, $\alpha_k$. However, solving this maximization problem is challenging due to the exhaustive search over the continuous interval $[-\pi,\pi)$. To address this complexity, we propose the design of a finite codebook by quantizing the angular domain as follows:
\begin{equation}
    \Ac_Q = \{\av(\Delta_1),\av(\Delta_2),\dots,\av(\Delta_Q)\},\label{eq:codebook}
\end{equation} where $Q$ is the quantization level and 
\begin{equation}
    \Delta_q = -\pi+\frac{2(q-1)}{Q}\pi,\;\; q \in [Q].
\end{equation} Nonetheless, it is tricky to directly seek an optimal solution using the aforementioned codebook $\Ac_Q$ due to the following limitations: i) the necessity of estimating the complex channel gain $\alpha_k$; ii) the inability to precisely measure the SINR for a given array response vector, as the received signal vector comprises both desired and interference signals. To address these limitations, the channel estimation process is divided into two parts (refer to Section~\ref{subsec:channel1} for further details):
\begin{enumerate}
    \item { Reconsidering the symbol vector $\sv\in\CC^{N\times 1}$ in \eqref{eq:transmissionvector}, the BS sends two pilot symbol vectors in $\sv(N-1)$ and $\sv(N)$, respectively. }Given a quantized array response vector $\av({\Delta_q})$, we estimate the complex channel gain $\alpha_k$ using the pilot symbol $\sv(N-1)$.
    \item From the estimated channel gain $\hat{\alpha}_{[k,q]}$, we evaluate the associated SINR and subsequently identify the optimal array response vector using the pilot symbol $\sv(N)$.
\end{enumerate} As only two pilot symbols are adequate for the accurate estimation of  downlink channels in the proposed CIRCLE method, the BS can effectively serve $K=N-2$ devices. Consequently, in dense IoT networks supported by massive MIMO, the pilot overhead becomes negligible as the ratio $(N-2)/N$ approaches 1. For the sake of brevity in describing the proposed channel estimation algorithm, nonetheless, we will disregard the impact of additive noise. However, these effects will be thoroughly investigated in our simulations.

\subsection{The Proposed Channel Estimation Algorithm}\label{subsec:channel1}
We first delineate how to estimate the complex channel gain $\alpha_k$. Assuming that $\av(\Delta_q)$ closely approximates the actual array response vector $\av(\theta_k)$, we use the linear combiner $\Fm_{N-1}$ corresponding to the $\sv(N-1)$ and the combined signal with the $\av(\Delta_q)$ is well-approximated as follows:
\begin{align}
    d(\av(\Delta_q),{\Fm}_{N-1},\yv_k) &\approx d(\av(\theta_k),{\Fm}_{N-1},\yv_k)\nonumber\\
    &\stackrel{(a)}{=} ({\bf 1}_N\oslash(\av(\theta_k))^{*})^{\transp}\Fm_{N-1}^{*}\yv_k \nonumber\\
    &\stackrel{(b)}{=}\sqrt{p_{\rm t}N}\alpha_k^{*}\sv(N-1),\label{eq:pilot1}
\end{align} where (a) comes from the definition of the proposed linear combiner in \eqref{eq:linearreceiver} and (b) comes from the Lemma 2. From \eqref{eq:pilot1} and utilizing the known pilot symbol $\sv(N-1)$, we can estimate the complex channel gain associated with the $q$-th quantized array response vector as follows:
\begin{equation}
    \hat{\alpha}_{[k,q]}^{*} = \frac{1}{\sqrt{p_{\rm t}N}\sv(N-1)}d(\av(\Delta_q),{\Fm}_{N-1},\yv_k).\label{eq:complexchannelgain}
\end{equation} It is important to note that the above estimator is well performed when the approximation in \eqref{eq:pilot1} is valid. However, significant SINR degradation occurs for the quantized array response vectors that deviate from the true vector $\av({\theta_k})$. This degradation can assist in identifying the desired quantized array response vector, as discussed follows. Now, we can estimate the SINR using another pilot symbol $\sv(N)$ as follows:
\begin{equation}
    \hat{\gamma}(\hat{\alpha}_{[k,q]}\av(\Delta_q),{\Fm}_N,\yv_k) = \frac{\EE\left[|\hat{P}_{[k,q]}|^2\right]}{\EE\left[|\hat{I}_{[k,q]}|^2\right]}.\label{eq:calculatedSINR}
\end{equation} where 
\begin{align}
    \hat{P}_{[k,q]} &= \sqrt{p_{\rm t}N}{\sv}(N),\\
    \hat{I}_{[k,q]} &= d(\hat{\alpha}_{[k,q]}\av(\Delta_q),{\Fm}_N,\yv_k)-{ \hat{P}_{[k,q]}}.
\end{align} Herein, the estimated signal power, denoted as $|\hat{P}_{[k,q]}|^2$, is derived from the maximum signal power in \eqref{eq:maxSINR}. Furthermore, the estimated interference-plus-noise power, denoted as $|\hat{I}_{[k,q]}|^2$, is obtained by subtracting the estimated signal $\hat{P}_{[k,q]}$ from the combined signal $d(\hat{\alpha}_{[k,q]}\av(\Delta_q),{\Fm}_N,\yv_k)$. From the estimated SINR, the estimated spectral efficiency for the $k$-th device, when employing the proposed method with the $q$-th quantized array response vector, is computed as follows:
\begin{equation}
    \widehat{\Rc}_{[k,q]} = \log_2\left(1+\hat{\gamma}(\hat{\alpha}_{[k,q]}\av(\Delta_q),\tilde{\Fm}_N,\yv_k)\right).\label{eq:SE}
\end{equation} By leveraging this, we can estimate the complex channel gain and the array response vector by exploring the candidates from the codebook $\Ac_Q$ as follows:

\vspace{0.2cm}
\noindent{\bf Iterations.} We initiate the iteration process with the initialization of $q_k^{\star} = 1$ for $k \in [K]$. For every $q \in [Q]$, we proceed to update the optimal index $q_k^{\star}$ based on the spectral efficiency:
\begin{equation}
    q_k^{\star} \leftarrow \argmax\left(\widehat{\Rc}_{[k,q]},\widehat{\Rc}_{[k,{q_k^{\star}]}}\right).\label{eq:comparison}
\end{equation} This process is summarized in Algorithm 1. Finally, the estimated channel for the device $k$ is obtained as:
\begin{equation}
    \hat{\hv}_k = \hat{\alpha}_{[k,q_k^{\star}]}\av(\Delta_{q_k^{\star}}).
\end{equation}

\begin{algorithm}[h]
\caption{The Proposed Channel Estimation Algorithm}
\begin{algorithmic}[1]

\State {\bf Input:} Received signal vectors $\{\yv_k:k\in[K]\}$ in \eqref{eq:concrsignal}, CSIT-Free precoding matrices $\{\tilde{\Fm}_n:n\in[N]\}$ in \eqref{eq:precoding}), and the quantized codebook with resolution $Q$, denoted as $\Ac_Q$ in \eqref{eq:codebook}

\State {\bf Initialization:} $q = 1$ and $q_k^{\star}=1,\;\forall k\in[K]$ 


\vspace{0.1cm}
\State {\bf Repeat until:} $q=Q$.

\begin{itemize}
    \item Estimate the complex channel gain via \eqref{eq:complexchannelgain}
    \item Estimate the spectral efficiency from  \eqref{eq:SE} and update $q_k^{\star}$ via \eqref{eq:comparison} 
    \item Set $q=q+1$
\end{itemize}


\vspace{0.1cm}
\State {\bf Output:} The indices of the optimal array response vectors $\{q_k^{\star}, \;\forall k\in[K]\}$
\end{algorithmic}
\end{algorithm}

\subsection{Robust Channel Estimation Utilizing OFDM} \label{subsec:wideband}
We further enhance the performance of the proposed channel estimation method by leveraging the abundant bandwidth resources avaliable in mmWave communications.
To be specific, we can significantly improve the robustness of the proposed method to additive noise and NLoS paths by jointly optimizing the array response vector while harnessing the frequency-independent channel parameters, such as the AoDs $\{\theta_k:k\in[K]\}$, in a wideband system. Before describing the proposed channel estimation method, we first explain the wideband massive MIMO systems. To mitigate the inter-symbol interference inherent in wideband communications, orthogonal frequency division multiplexing (OFDM) is employed. In this approach, the signal is generated in the frequency domain and then transformed into the time domain by inverse discrete Fourier transform (IDFT). Let $M$ denote the number of subcarriers and $L_{\rm CP}$ the length of the cyclic prefix in OFDM, while $B$ represents the system bandwidth. Consequently, the frequency of subcarrier $m\in[M]$ is determined as $f_m = f_{\rm c} + B(2m-1-M)/(2M)$. 

Considering the OFDM systems, the received signal on the $m$-th subcarrier can be expressed as follows:
\begin{equation}
    \yv_{[m,k]} = \sqrt{p_{\rm t}}\hv_{[m,k]}^{\herm}\xv_m + z_{[m,k]},
\end{equation} where the DL channel response $\hv_{[m,k]}\in\CC^{N\times 1}$ is given by
\begin{equation}
    \hv_{[m,k]} = \alpha_{[m,k]}\av_m(\theta_k), 
\end{equation} and where $\xv_m\in\CC^{N\times 1}$ denotes the $n$-th DL transmit vector on the $m$-th subcarrier, and $z_{[m,k]}\sim \Cc\Nc(0,\sigma^{2})$ represents the additive white Gaussian noise. For $\lambda_m = c/f_m$, the array response vector on the $m$-th subcarrier, denoted by $\av(\theta,f_m)$, is defined as follows:
\begin{equation}
    \av_m(\theta) = \left[1,e^{j\pi\frac{\lambda_{\rm c}}{\lambda_m}\sin{\theta}},\dots,e^{j \pi\frac{\lambda_{\rm c}}{\lambda_m}(N-1)\sin{\theta}}\right]^{\transp}.\label{eq:OFDM_a}
\end{equation} Here, we assume that the discrete Fourier transform (DFT) operation and cyclic prefix removal have been conducted at the devices. We refer to \cite{Wang2018wideband} for the detailed derivation of the frequency-domain channels in OFDM systems. For the $m$-th subcarrier, the concatenation of the received signals at the $k$-th device is represented as follows:
\begin{equation}
    \yv_{[m,k]} = \begin{bmatrix}
        y_{[1,m,k]} &\cdots & y_{[N,m,k]}
    \end{bmatrix}^{\transp}\in\CC^{N\times 1}.\label{eq:Mconcrsignal}
\end{equation} Herein, the received signal at the $n$-th time slot is expressed as follows:
\begin{equation}
    y_{[n,m,k]} = \sqrt{p_{\rm t}}\hv_{[m,k]}^{\herm}\xv_{[n,m]} + z_{[n,m,k]},
\end{equation} where $z_{[n,m,k]}\sim\Cc\Nc(0,\sigma^{2})$. As in Section~\ref{subsec:precoding}, we construct the transmit vector for the $n$-th time slot on the $m$-th subcarrier as follows:
\begin{equation}
    \xv_{[n,m]} = \Pm_n\sv_m,
\end{equation} where $\sv_m\in\CC^{K\times 1}$ represents the symbol vector on the $m$-th subcarrier, including two pilot symbols in $\sv_m([N-1:N])$. Noticeably, we employ the identical precoding design as specified in \eqref{eq:precoding} for all subcarriers. Since $\theta_k$ is frequency-independent, we can significantly enhance the channel estimation accuracy by jointly optimizing the array response vectors across the $M$ subcarriers. To this end, we first design the codebook for the $m$-th subcarrier,  similar to that in \eqref{eq:codebook}:
\begin{equation}
    \Ac_{[m,Q]} = \{\av_m(\Delta_1),\av_m(\Delta_2),\dots,\av_m(\Delta_Q)\},\label{eq:Mcodebook}
\end{equation} where $\av_m(\cdot)$ is defined in \eqref{eq:OFDM_a}. As performed in \eqref{eq:complexchannelgain}, we can estimate the complex channel gain for the $m$-th subcarrier as follows:
\begin{equation}
    \hat{\alpha}_{[m,k,q]}^{*} = \frac{1}{\sqrt{p_{\rm t}N}\sv_{m}(N-1)}d(\av_m(\Delta_q),{\Fm}_{N-1},\yv_{[m,k]}).\label{eq:Mcomplexchannelgain}
\end{equation} From \eqref{eq:calculatedSINR} and using the estimated complex channel gain, we can estimate the SINR for the $m$-th subcarrier:
\begin{equation}
    \hat{\gamma}\left(\hat{\alpha}_{[m,k,q]}\av_m(\Delta_q),{\Fm}_N,\yv_{[m,k]}\right) = \frac{\EE\left[|\hat{P}_{[m,k,q]}|^2\right]}{\EE\left[|\hat{I}_{[m,k,q]}|^2\right]},\label{eq:McalculatedSINR}
\end{equation} where  
\begin{align}
    \hat{P}_{[m,k,q]} &= \sqrt{p_{\rm t}N}{\sv}_{m}(N),\\
    \hat{I}_{[m,k,q]} &= d(\hat{\alpha}_{[m,k,q]}\av_m(\Delta_q),{\Fm}_N,\yv_{[m,k]})-\hat{P}_{[m,k,q]}.
\end{align} Leveraging this SINR, we can derive the estimated spectral efficiency for the $m$-th subcarrier at the $k$-th device is computed as follows:
\begin{equation}
    \widehat{\Rc}_{[m,k,q]} = \log_2\left(1+\hat{\gamma}(\hat{\alpha}_{[m,k,q]}\av_m(\Delta_q),{\Fm}_N,\yv_{[m,k]})\right).
\end{equation} By taking the average over the $M$ subcarriers, we can obtain the mean-spectral efficiency as follows:
\begin{equation}
    \bar{\Rc}_{[k,q]} = \frac{1}{M+L_{\rm CP}}\sum_{m=1}^{M}\widehat{\Rc}_{[m,k,q]}.\label{eq:meanSE}
\end{equation} Notably, since the physical AoD, denoted as $\theta_k$, is frequency-independent, we can estimate the AoD using the mean-spectral efficiency in \eqref{eq:meanSE}. This approach can enhance the robustness of the proposed method against additive noise and NLoS paths.

\begin{algorithm}[h]
\caption{The Robust Channel Estimation Algorithm}
\begin{algorithmic}[1]

\State {\bf Input:} Received signal vectors $\{\yv_{[m,k]}:m\in[M],k\in[K]\}$ in \eqref{eq:Mconcrsignal}, CSIT-Free precoding matrices  $\{{\Fm}_n:n\in[N]\}$ in \eqref{eq:precoding}, and the quantized codebooks with resolution $Q$, denoted as $\{\Ac_{[m,Q]}:m\in[M]\}$ in \eqref{eq:Mcodebook}.

\State {\bf Initialization:} $q = 1$ and  $q_k^{\star}=1,\;k\in[K]$ 
\vspace{0.1cm}
\State {\bf Repeat until:} $q=Q$.

\begin{itemize}
    \item Estimate the complex channel gain for the $m$-th subcarrier via \eqref{eq:Mcomplexchannelgain}
    \item Estimate the spectral efficiency from \eqref{eq:comparison} and update $q_k^{\star}$ via \eqref{eq:comparisonOFDM}. 
    \item Set $q=q+1$
\end{itemize}

\vspace{0.1cm}
\State {\bf Output:} The indices of the optimal array response vectors $\{q_k^{\star},\;k\in[K]\}$

\end{algorithmic}
\end{algorithm}

\vspace{0.2cm}
\noindent{\bf Iterations.} We commence the iteration process with the initialization of $q_k^{\star} = 1$ for $k \in [K]$. For every $q \in [Q]$, we proceed to update the optimal index $q_k^{\star}$ based on the mean-spectral efficiency:
\begin{equation}
    q_k^{\star} \leftarrow \argmax(\bar{\Rc}_{[k,q]},\bar{\Rc}_{[k,{q_k^{\star}]}}).\label{eq:comparisonOFDM}
\end{equation} 
This process is summarized in Algorithm 2. Finally, the estimated channel for the $m$-th subcarrier at the $k$-th device is obtained as:
\begin{equation}
    \hat{\hv}_{[m,k]} = \alpha_{[m,k,q_k^{\star}]}\av_m(\Delta_{q_k^{\star}}).
\end{equation}

\section{Analysis}

In this section, we perform a comprehensive analysis of the performance of our CIRCLE method in practical applications. We refer to the CIRCLE method when integrated with the robust channel estimation algorithm in Algorithm 2 for wideband OFDM systems, as Robust-CIRCLE ({\bf R-CIRCLE}). We compare the latency of the R-CIRCLE method with existing CSIT-Based methods. We begin by presenting the computational complexity of the proposed R-CIRCLE method. Subsequently, we investigate the performance bounds of our method, ultimately revealing the principle findings of this paper.

\subsection{Complexity and Latency Analysis}
To analyze the computational complexity of the proposed R-CIRCLE method, we evaluate the number of complex multiplications required by the proposed linear combiner. This metric has been employed in the literature to assess the complexity of MIMO algorithms \cite{Chen2019,Lee2024near}. 
Given a quantized array response vector on a subcarrier, denoted as $\av_m(\Delta_{q})$, the operation of the linear combiner is described in \eqref{eq:linearreceiver}. To avoid duplicate multiplications, we first calculate the products of $\Fm_N$ and $\yv_{[m,k]}$, and $\Fm_{N-1}$ and $\yv_{[m,k]}$ for channel estimation, each of which requires $N^2$ complex multiplications. Additionally, $N$ complex multiplications are needed for the product involving the quantized array response vector. Consequently, the total number of complex multiplications for operating the R-CIRCLE method is computed as follows:
\begin{equation}
    \psi = M(2N^2 + Q N).
\end{equation}

\begin{figure}[!t]
    \centering
    \subfigure[The conventional CSIT-Based method.]{
         \includegraphics[width=0.7\linewidth]{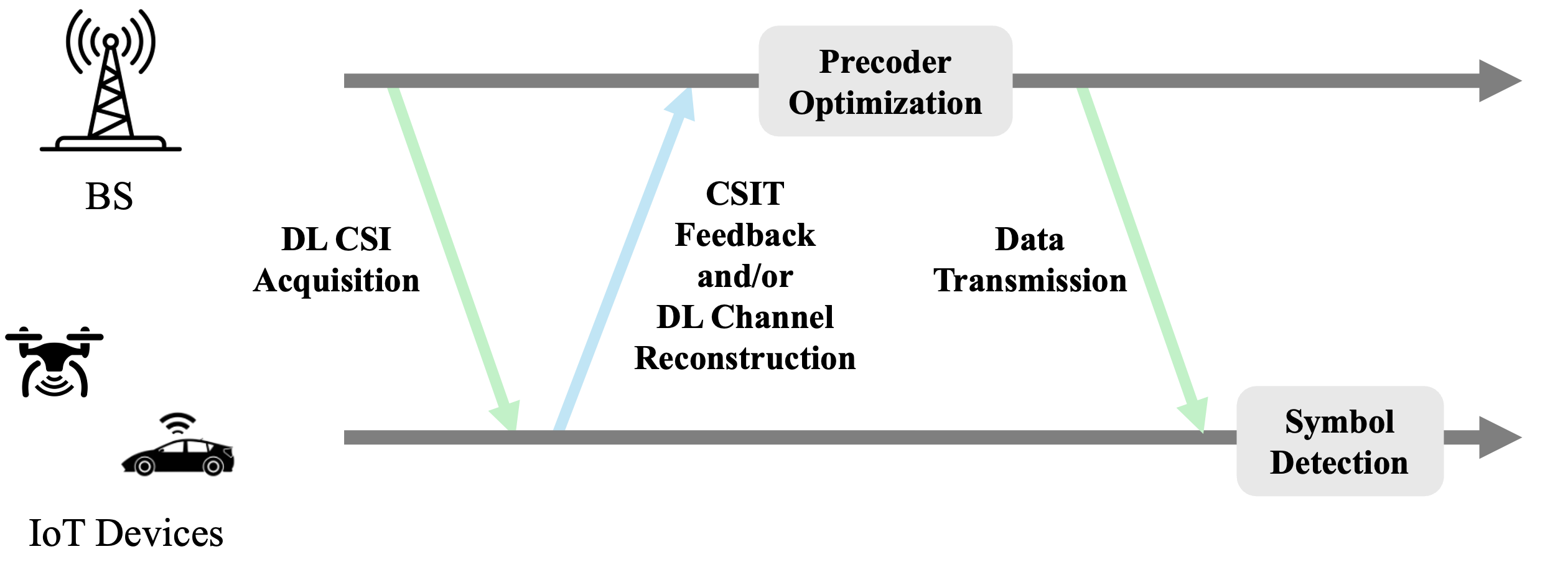}}
    \subfigure[The proposed CSIT-Free method.]{\includegraphics[width=0.47\linewidth]{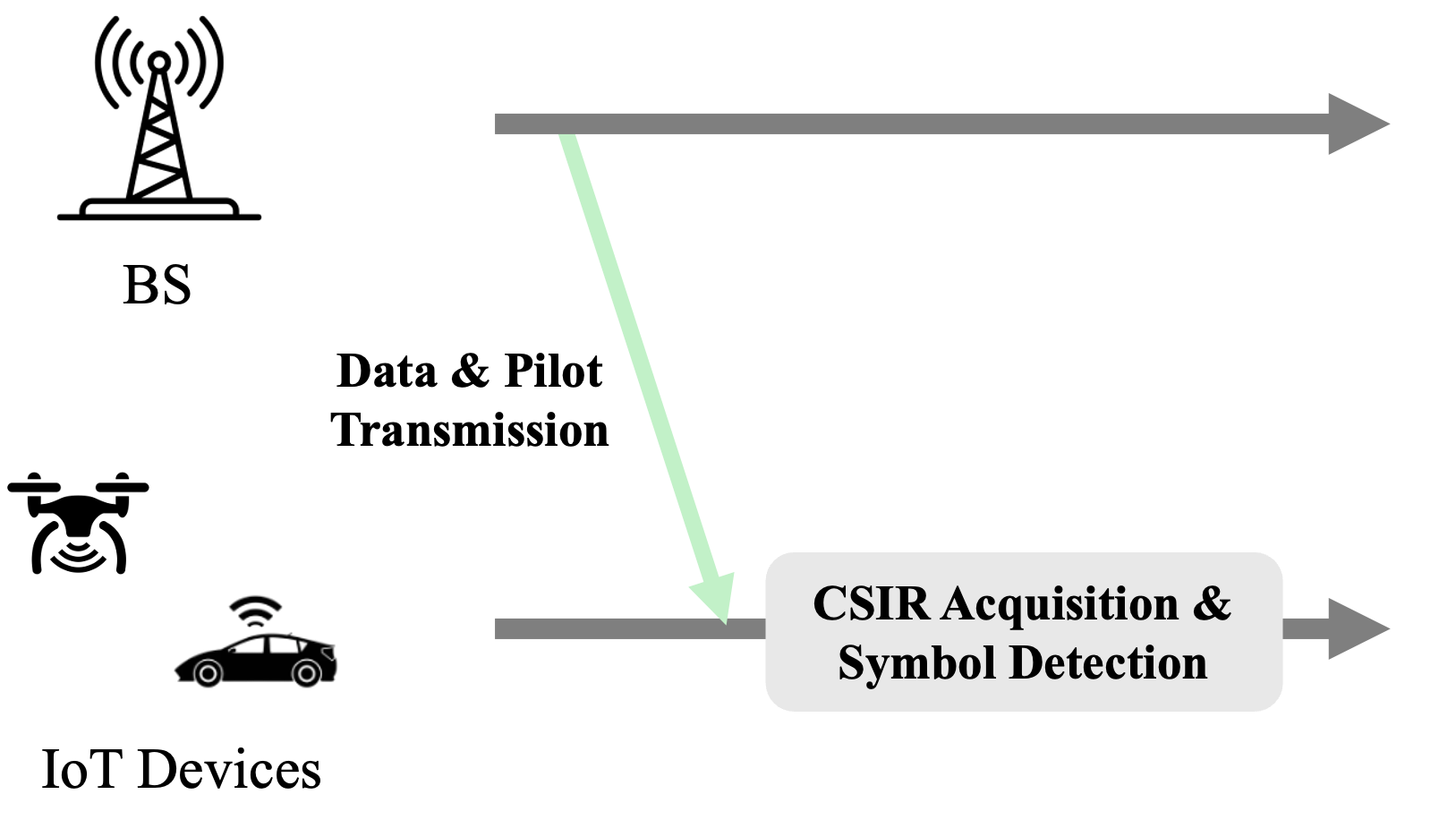}}
    \caption{Description of the downlink data transmission process.}
    \label{fig:simulation_graph}
\end{figure}

We can categorize downlink transmission into two types: { i) CSIT-Based methods and ii) CSIT-Free methods.} As depicted in Fig. 3, the CSIT-Based method necessitates the acquisition of downlink CSI, CSIT feedback (or downlink channel reconstruction from the uplink channel, which eliminates the need for a separate downlink CSI acquisition process or significantly reduces the amount of the CSI feedback), precoding optimization, data transmission, and symbol detection. This sequence of uplink channel estimation, downlink channel reconstruction, and precoding optimization incurs significant overhead and computational complexity, particularly as the number of antennas and devices increases in IoT networks. Conversely, in the CSIT-Free approach, such as the proposed CIRCLE method, the BS simply transmits data and pilot symbols to the IoT devices, enabling each device to individually acquire CSIR and detect the desired symbol using the proposed channel estimation and precoding methods. Although the overhead and complexity of the CIRCLE method also increase with the number of devices, they remain comparable to, or even lower than, those of the CSIT-Based methods. Furthermore, the interference-free nature of the CIRCLE method facilitates the achievement of high spectral efficiency with a relatively small number of antennas. In contrast, the CSIT-Based methods typically require a significantly larger number of antennas due to their reliance on suboptimal precoding methods that aim to reduce latency and complexity of precoding optimization. In our simulations, we will demonstrate the superiority and practicality of our CIRCLE method in comparison to existing CSIT-Based benchmark methods.

\subsection{Performance Analysis}
From the maximum SINR in \eqref{eq:maxSINR} with full-CSIR, we can derive the maximum spectral efficiency of the $k$-th device on the $m$-th subcarrier in wideband OFDM systems as follows:
\begin{align}
    \Rc_{[m,k]}^{\star} 
     &\eqdef \log_{2}\left(1+\frac{p_{\rm t}{\hv}_{[m,k]}^{\transp}{\hv}_{[m,k]}^{*}}{\sigma^2}\right).
\end{align} Subsequently, the maximum sum-spectral efficiency in a wideband OFDM system is defined as follows:
\begin{equation}
\Wc^{\star}=\frac{1}{M+L_{\rm CP}}\sum_{k=1}^{K}\sum_{m=1}^{M}\Rc_{[m,k]}^{\star}.\label{eq:maxSR}
\end{equation} From \eqref{eq:McalculatedSINR}, the achievable sum-spectral efficiency of the $k$-th device on the $m$-th subcarrier of the proposed R-CIRCLE method, when the proposed robust channel estimation algorithm in Algorithm 2 is applied, is defined as follows:
\begin{equation}
    {\Rc}_{[m,k]}^{\rm prop} = \log_2\left(1+\gamma(\hat{\hv}_{[m,k]},\Fm_k, \yv_{[m,k]})\right).
\end{equation} Accordingly, the achievable sum-spectral efficiency of the proposed method in wideband OFDM systems is computed as follows:
\begin{equation}
    {\Wc}^{\rm prop} = {\frac{1}{M+L_{\rm CP}}}\sum_{k=1}^{K}\sum_{m=1}^{M}\Rc^{\rm prop}_{[m,k]}.\label{eq:estimatedSR}
\end{equation} Now, we are now ready to state the main result of this paper.
\vspace{0.1cm}

\begin{theorem}
When the LoS channel is dominant, as the transmit power and the quantization resolution become large, the proposed R-CIRCLE method nearly achieves the maximum sum-spectral efficiency:
    \begin{equation}
        \lim_{p_{\rm t},Q\rightarrow\infty}{\Wc^{\rm prop}} = \Wc^{\star}.
    \end{equation}
\end{theorem}
\begin{IEEEproof}
    From corollary 1, it is evident that the estimation accuracy of the complex channel gain and the array response vector is enhanced as the noise effect decreases and the quantization resolution increases. As the transmit power and the quantization resolution become large, therefore, the achievable sum-spectral efficiency of the R-CIRCLE method nearly converges to the maximum sum-spectral efficiency. This completes the proof.
\end{IEEEproof} 

\vspace{0.2cm}
\noindent In our simulations, we will demonstrate the validity of our analysis presented in Theorem 1, even when considering finite and practical quantization resolution along with practical SNRs. Consequently, the proposed method performs effectively while maintaining affordable computational complexity and power consumption. Furthermore, despite the presence of additional NLoS paths, we will verify that Theorem 1 is valid based on the experimental analysis in Fig. 2.

\begin{figure*}[!t]
    \centering
    \subfigure[$\rho = 1/32$]{
        \includegraphics[width=0.43\linewidth]{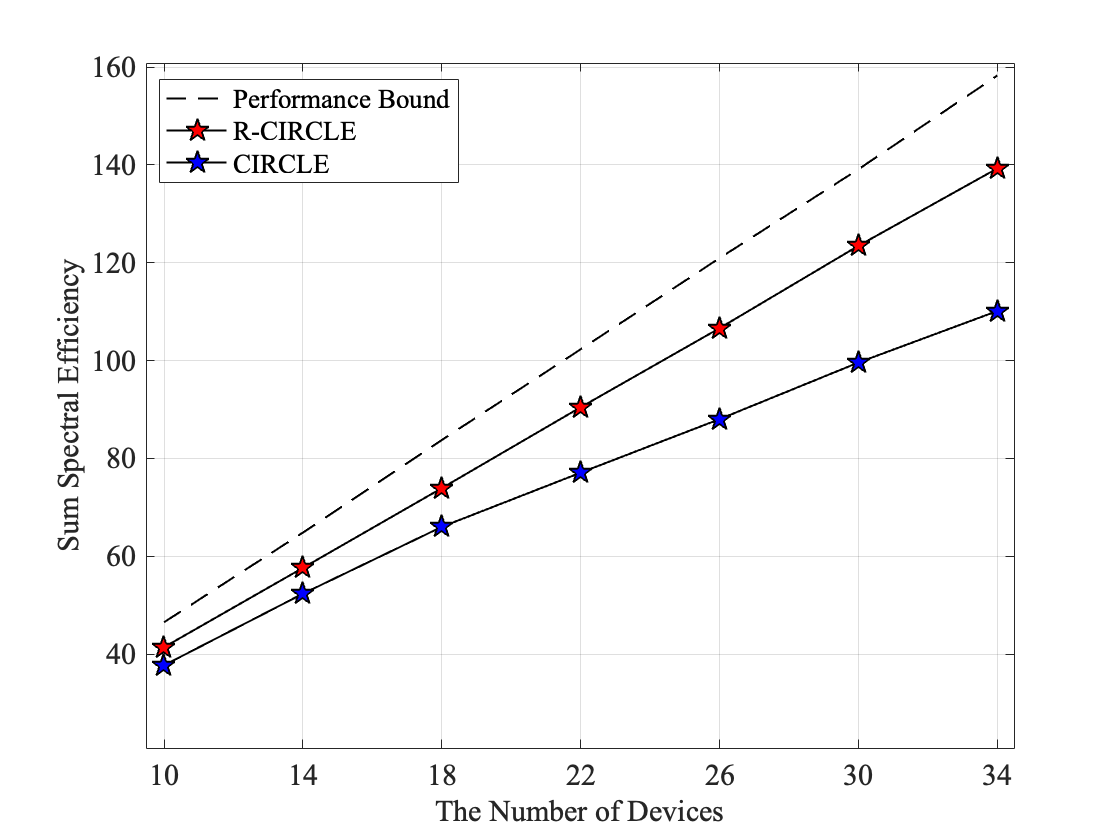}
    }
    \subfigure[$\rho = 1/8$]{
        \includegraphics[width=0.43\linewidth]{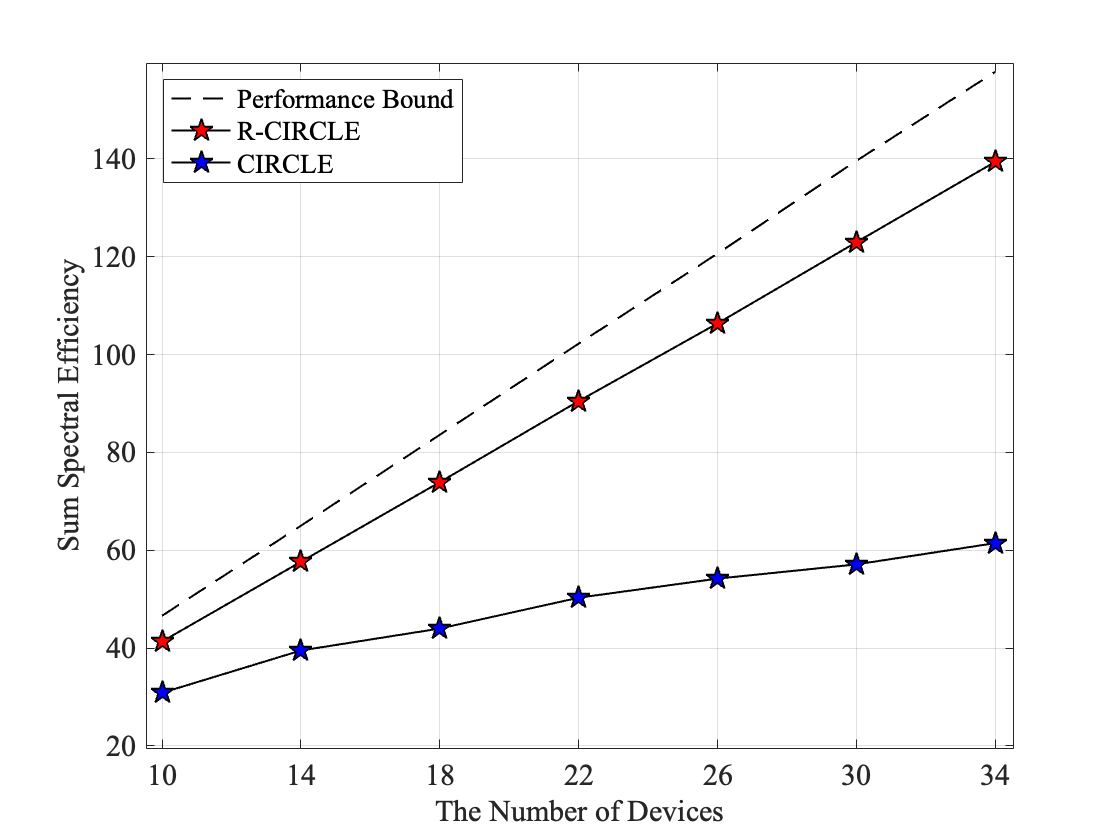}
    }
    \subfigure[$\rho = 1/2$]{
        \includegraphics[width=0.43\linewidth]{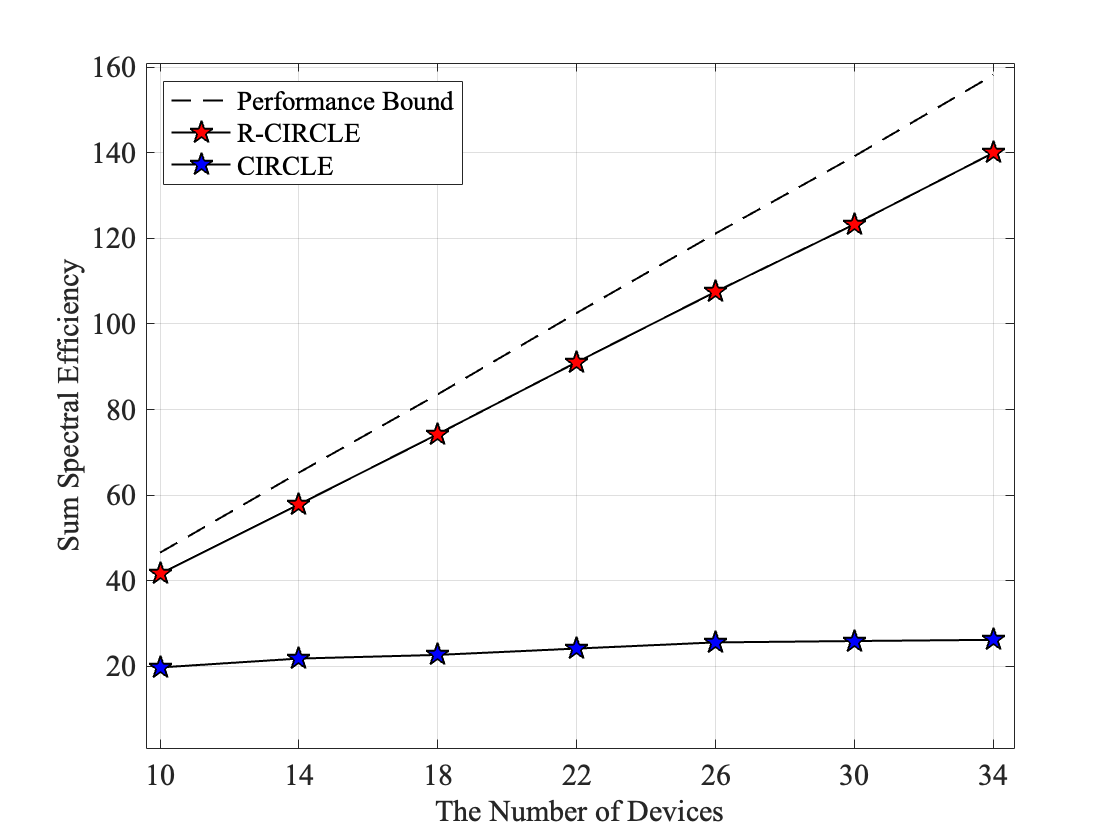}
    }
    \subfigure[$\rho = 2$]{
        \includegraphics[width=0.43\linewidth]{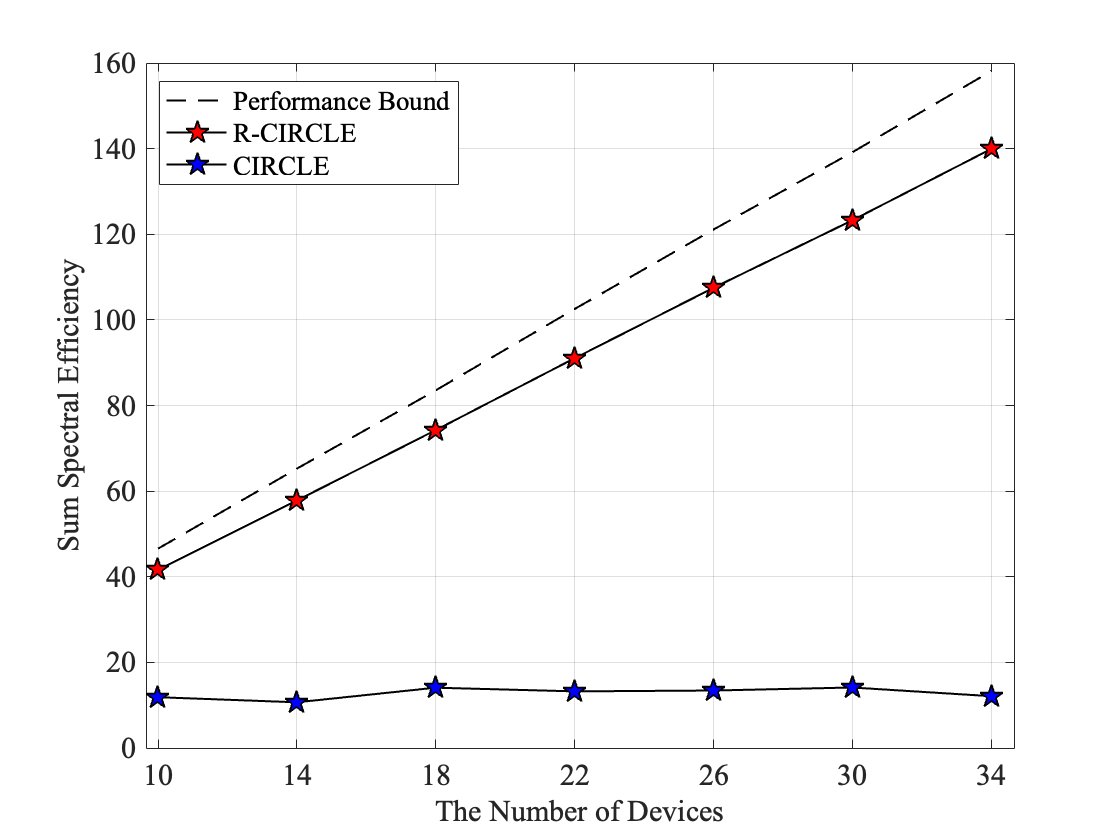}
    }
    \caption{The sum-spectral efficiency as a function of the number of devices. $\SNR=10$ dB. } \label{fig:action_tabular}
\end{figure*}

\section{Simulation Results}

In this section, we provide the simulation results to verify the effectiveness of the proposed CIRCLE method. The parameters for this analysis are as follows:
$f_{\rm c} = 100$ GHz, $B = 10$ GHz, $M = 10$, $L_{\rm CP} = 4$ and $L_k = 3$ for all $k \in [K]$. The signal-to-noise ratio (SNR) of the received signal at the $k$-th device on the $m$-th subcarrier is defined as follows:
\begin{equation}
    \SNR = 10\log_{10}\left(\frac{\EE[|\sqrt{p_{\rm t}}\hv_{[k,m]}\xv_m|^2]}{\sigma^2}\right) \mbox{ dB},
\end{equation} where  $\sigma^2 = -10$ dB and for all $k \in [K]$ and $m \in [M]$, the complex channel gains for the LoS path and the NLoS paths are respectively given by $\alpha_{[k,m]}\sim\Cc\Nc(0,1)$ and $\beta_{[k,m]}\sim\Cc\Nc(0,\delta^2)$ with $\delta^2 = -15$ dB. We consider the following benchmark schemes:
\begin{itemize}
    \item {\bf Performance Bound}: This indicates the maximum sum-spectral efficiency achieved under full-CSIR, defined as $\Wc^{\star}$ in \eqref{eq:maxSR}, which serves as the performance bound for the proposed CIRCLE method. 
    \item {\bf CIRCLE}: In OFDM systems, the CIRCLE method estimates the channels among the subcarriers separately via Algorithm 1.  In contrast, the R-CIRCLE method estimates the channels jointly across the subcarriers via Algorithm 2. The comparative analysis of the CIRCLE and R-CIRCLE methods substantiates the effectiveness of the proposed robust channel algorithm in Algorithm 2.
    \item {\bf WMMSE}: Under the assumption of full-CSIT, the BS performs the weighted minimum mean square error (WMMSE) precoding.
    \item {\bf ZF}: Under the assumption of full-CSIT, the BS performs the zero-forcing (ZF) precoding.
    \item {\bf W/O CSIT Feedback \cite{Han2024}}: As the best-known work to reduce the channel estimation overhead in FDD massive MIMO, this method reconstructs the CSIT by utilizing the uplink pilots without CSIT feedback, under the assumption that the uplink channels are perfectly estimated. Subsequently, the information symbols are precoded using the generalized power iteration precoding (GPIP) method based on the reconstructed downlink channels.
    \item {\bf MRT}: Under the assumption of full-CSIT, this method employs the maximum-ratio-transmission (MRT) precoding for information symbols.
\end{itemize} 
Regarding the CSIT-Based benchmark methods, the received signal at the $k$-th device on the $m$-th subcarrier is expressed as follows:
\begin{equation}
    y_{[m,k]}^{\rm CSIT} = \frac{N}{K}\sqrt{p_{\rm t}}\hv_{[m,k]}^{\herm}\left(\xv_{[m,k]}^{\rm CSIT} + \sum_{k'\neq k}^{K}\xv_{[m,k']}^{\rm CSIT}\right) + z_{[m,k]}^{\rm CSIT},\label{eq:receivedCSIT}
\end{equation} where the transmit power constraint is defined as $\left|\xv_{[m,k]}^{\rm CSIT}\right|_2^2 = 1$. Additionally, the downlink transmission vector is defined as follows:
\begin{equation}
     \xv_{[m,k]}^{\rm CSIT} = \sv(k)\fv_{[m,k]}^{\rm CSIT}\in\CC^{N\times 1},
\end{equation} where $\fv_{[m,k]}^{\rm CSIT}\in\CC^{N\times 1}$ represents the CSIT-Based precoding vector, which is fully determined by the selected precoding method (e.g., WMMSE, ZF, and MRT). Remarkably, the multiplicative term $N/K$ (i.e., the power normalization) is introduced to facilitate a fair comparison with the R-CIRCLE method, accounting for the impact of pilot symbols utilized in channel estimation.
Subsequently, the sum-spectral efficiency of the CSIT-Based benchmark methods is defined as follows:
\begin{equation}
    \Wc^{\rm CSIT}  = \frac{1}{M+L_{\rm CP}}\sum_{k=1}^{K}\sum_{m=1}^{M}\Rc_{[m,k]}^{\rm CSIT},
\end{equation} where 
\begin{equation*}
    \Rc_{[m,k]}^{\rm CSIT} = \log_2\left(1+\frac{\left|\frac{N}{K}\sqrt{p_{\rm t}}\hv_{[m,k]}^{\herm}\xv_{[m,k]}^{\rm CSIT}\right|^2}{\sum_{k'\neq k}^{K}\left|\frac{N}{K}\sqrt{p_{\rm t}}\hv_{[m,k]}^{\herm}\xv_{[m,k']}^{\rm CSIT}\right|^2 + \sigma^2}\right).
\end{equation*} 
{For the CIRCLE and R-CIRCLE methods,} the quantization resolution is set to $Q=512$ in accordance with the UL channel estimation algorithms that utilize the quantization codebook \cite{tsai2018efficient, chen2023channel}. The expectation of the sum-spectral efficiency is evaluated by Monte Carlo simulations, conducted with $10^3$ trials.


\begin{figure}[t]
\centering
\includegraphics[width=0.5\linewidth]{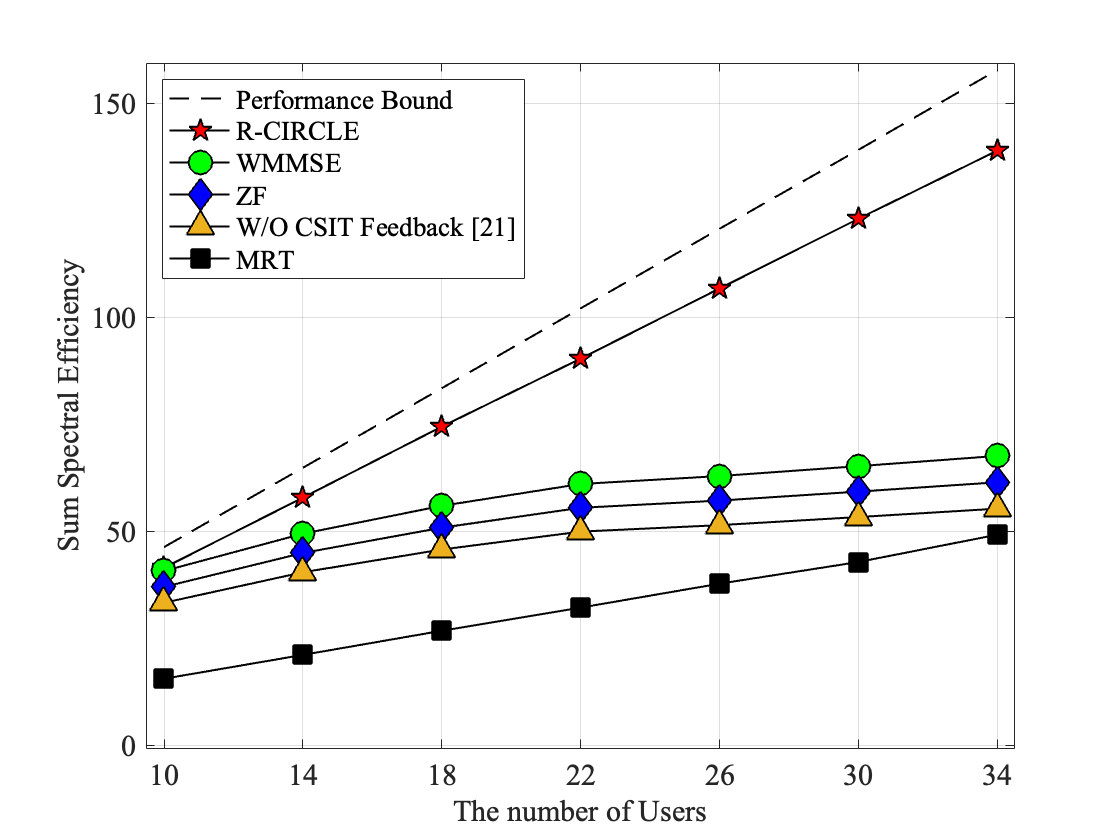}
{ \caption{The performance comparison of various methods as a function of the number of IoT devices, denoted as $K$.
$\SNR=10$ dB and $\rho = 2$ (i.e., $\theta_k \in [0,2\pi)$).} } 
\end{figure}

\begin{figure}[t]
\centering
\includegraphics[width=0.5\linewidth]{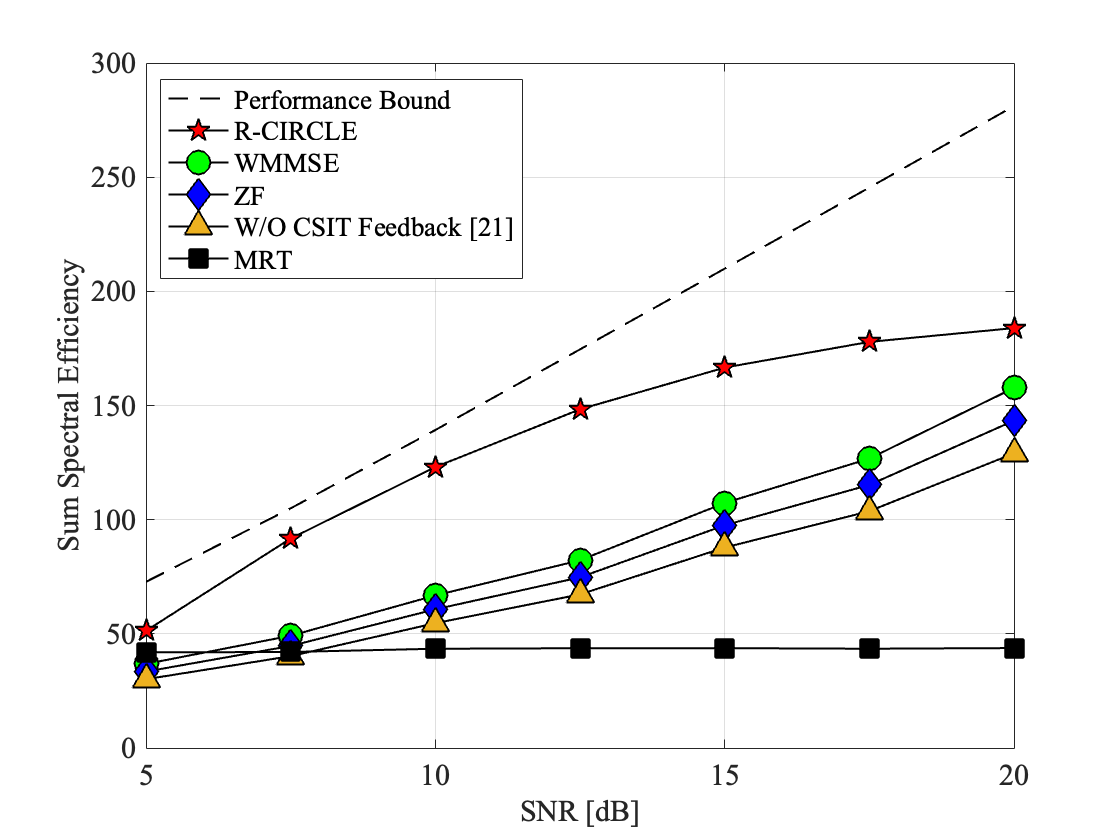}
\caption{The performance comparison of various methods as a function of SNR. $K=30$ and $\rho = 2$ (i.e., $\theta_k \in [0,2\pi)$).}
\end{figure}

Fig. 4 depicts the sum-spectral efficiency as a function of the number of IoT devices, where the angular domain for the LoS channel is determined by the range parameter $\rho$ as follows: $\theta_k\in[0,\rho\pi)$. Thus, the density of the devices can be controlled using the range parameter $\rho$. Additionally, the codebooks $\{\Ac_{[m,Q]}:m\in[M]\}$ are meticulously designed to account for the specified angular domain. It is observed that the CIRCLE method shows significant performance degradation, particularly as the range parameter increases. Within a limited angular region, such as $\rho = 1/32$ and $1/8$ in Fig. 4(a) and (b), the sum-spectral efficiency of the CIRCLE method improves with the increasing number of devices, although it remains considerably lower than the performance bound. When the devices are located in a wider angular domain, specifically $\rho = 1/2$ and $2$ in Fig. 4(c) and (d), the sum-special efficiency of the CIRCLE method fails to exhibit any performance gain despite the increasing number of devices. This phenomenon can be attributed to the widening angular domain, which results in a larger quantization error, thereby complicating the identification of a suitable array response vector from the codebook. In contrast, the proposed R-CIRCLE method consistently demonstrates superior performance compared to the CIRCLE method, irrespective of the range parameter $\rho$. This finding implies that the joint optimization approach, as indicated by the mean of the the spectral efficiency in \eqref{eq:meanSE}, effectively identifies the optimal array response vector even in the presence of an expanded angular domain. Furthermore, under practical quantization resolution and SNRs in conjunction with the NLoS paths, the R-CIRCLE method closely approaches the performance bound, thereby validating the practical applicability of our theoretical analysis.


Fig. 5 illustrates the sum-spectral efficiency as a function of the number of devices for various methods. Notably, the sum-spectral efficiency of both the performance bound and the proposed R-CIRCLE method exhibits a linear increase with the number of devices, indicating that the proposed method effectively cancels interference signals, even as the number of devices increases. {The performance gap of the R-CIRCLE method relative to the performance bound is attributed to pilot overhead for channel estimation, as indicated by the relationship $K=N-2$. In practical scenarios, this gap appears to be inevitable, as the performance bound is based on the unrealistic assumption of perfect CSI.} In contrast, the sum-spectral efficiency of the benchmark methods tend to saturate as the number of devices increases, despite the full availability of CSIT. Notably, the CSIT reconstruction method, referred to as ``W/O CSIT Feedback \cite{Kim2024}'', experiences performance degradation due to the reconstruction errors. Since the CSIT reconstruction method presumes that the uplink channels are perfectly estimated, it is anticipated that the performance of this method will further diminish under realistic conditions. Additionally, as indicated in \eqref{eq:receivedCSIT}, it is assumed that the transmit power of the CSIT-Based benchmark methods is equivalent to the total power consumption of the proposed method. However, the CSIT-Based methods require additional power consumption to acquire CSIT, which is not accounted for in our analysis. Consequently, we can conclude that the proposed method significantly outperforms the CSIT-Based benchmark methods, even with lower energy expenditure, particularly in dense IoT networks.


Fig. 6 illustrates the sum-spectral efficiency as a function of the SNR for various methods. Notably, the proposed R-CIRCLE method exhibits saturation as the SNR increases. This saturation occurs because, despite the enhanced SNR, the quantization error resulting from channel estimation causes an error floor, thereby limiting further performance improvement. Indeed, it is possible to mitigate the error floor by increasing the quantization resolution (i.e., $Q$); however, this approach incurs high computational complexity for channel estimation, which, in turn, leads to increased latency. Nonetheless, the proposed R-CIRCLE method consistently outperforms the CSIT-Based benchmark methods across all SNR regimes. In practical SNR regimes (e.g., from 5 dB to 10 dB), the performance gap between the proposed method and the CSIT-Based methods is particularly pronounced, exceeding that observed at higher SNRs. This finding suggests that the deployment of the proposed method offers practical advantages, demonstrating superior performance in low-to-intermediate SNR conditions compared to the CSIT-Based methods.


Our experimental analysis suggests that the proposed R-CIRCLE method is a strong candidate for dense and URLLC IoT networks, as it meets stringent requirements through its low latency and energy expenditure, resulting from its CSIT-Free nature. Additionally, the high spectral efficiency derived from the interference-free characteristics of the proposed method further supports its suitability for such applications.


\section{Conclusion}
We investigated the downlink data transmission problem in dense IoT networks, which necessitate URLLC, low energy expenditure, and high spectral efficiency. To efficiently support these networks, we considered FDD massive MIMO, which facilitates simultaneous UL and DL communications. 
To address the excessive overhead and power consumption associated with CSIT acquisition in FDD massive MIMO, we developed a CSIT-Free downlink data transmission strategy. In particular, we proposed the CSIT-Free precoding method, termed R-CIRCLE, which enables interference-free signal combining at the IoT devices. Through simulations, we validated the theoretical analysis of the proposed method in practice, and demonstrated its superiority compared to the conventional CSIT-Based methods.

\appendices
\section{Proof of Lemma 1}\label{subsec:lemma1}

We first define an $N\times N$ DFT matrix as follows: 
\begin{align}
    \Um_N = \begin{bmatrix}
        \uv_1 &\cdots & \uv_N
    \end{bmatrix},
\end{align} where 
\begin{equation}
    \uv_n = \left[
        (\omega^{n-1})^{0}, (\omega^{n-1})^{1}, \dots ,(\omega^{n-1})^{N-1}
    \right]^{\transp}\in\CC^{N\times 1},\label{eq:dftvec}
\end{equation} and $\omega = e^{-j(\frac{2\pi}{N})}$. Noticeably, it holds that $\sum_{n=1}^{N}\omega^{n-1} = 0$. From the circulant matrix $\Cm_N$ defined in \eqref{eq:circulant}, the $k$-th permutation matrix in \eqref{eq:permutation} can be represented by
\begin{align}
    \Um_{[N,k]} &= \Um_N\Pm_k\nonumber\\
    &=\begin{bmatrix}
        \uv_{\Cm_N(1,k)} &\cdots & \uv_{\Cm_N(N,k)}
    \end{bmatrix},
\end{align} where an $N\times N$ permutation matrix is given by
\begin{equation}
    \Pm_k = \begin{bmatrix}
        \Id_N(:,\Cm_N(1,k)) &\cdots &\Id_N(:,\Cm_N(N,k))
    \end{bmatrix},
\end{equation} satisfying $\Pm_k\Pm_k^{\herm} = \Id_N$. Then, it holds that 
\begin{equation}
    \Um_{[N,k]}\Um_{[N,k]}^{\herm} = \Um_N\Pm_k\Pm_k^{\herm}\Um_N^{\herm} = \Id_N.
\end{equation} For $k\neq k'$, on the other hand, we have:
\begin{align}
    \Um_{[N,k]}\Um_{[N,k']}^{\herm} = \sum_{i=1}^{N}\uv_{\Cm_N(i,k)}\uv_{\Cm_N(i,k')}^{\herm}.
\end{align} For each $i\in[N]$, the $(\alpha,\beta)$-th entry of $\uv_{\Cm_N(i,k)}\uv_{\Cm_N(i,k')}^{\herm}$ is given by
\begin{align}
    &(\omega^{\Cm_N(i,k)-1})^{\alpha-1}((\omega^{\Cm_N(i,k')-1})^{\beta-1})^{*}\nonumber\\
    &\quad =\omega^{((\Cm_N(i,k)-1)(\alpha-1) - (\Cm_N(i,k')-1)(\beta-1))}\nonumber\\
    &\quad =c_{[i,k,k']}\omega^{(\alpha\Cm_N(i,k) - \beta\Cm_N(i,k'))},
\end{align} where $c_{[i,k,k']} = \omega^{\beta-\alpha+\Cm_N(i,k')-\Cm_N(i,k)}\neq 0$. From these notations, we now have the $(\alpha,\beta)$-th entry of $\Um_{[N,k]}\Um_{[N,k']}^{\herm}$:
\begin{align}
    &\omega^{\beta-\alpha} \sum_{i=1}^{N}\omega^{(\alpha-1)\Cm_{N}(i,k)-(\beta-1)\Cm_{N}(i,k')}\nonumber\\
    &\quad =\omega^{\beta-\alpha}\omega^{\gamma(\beta-1)}\sum_{i=1}^{N}\omega^{(\alpha-\beta)\Cm_{N}(i,k)},\label{eq:sumN}
\end{align} where $\gamma=C_N(i,n)-C_N(i,n')\; \mbox{mod}\; N, \; \forall i \in [N]$, is a non-zero element belonging to $[N]$. When $\alpha=\beta$, the expression in \eqref{eq:sumN} is equal to
\begin{equation}
    N \omega^{\gamma(\alpha-1)} \neq 0.
\end{equation} Next, we consider the case of $\alpha\neq \beta$:
\begin{align}
   \sum_{i=1}^{Q}\omega^{(\alpha-\beta)\Cm_{Q}(i,q)} = 0,
\end{align} because this is equivalent to the sum of $i$-th column of the DFT matrix with $i\neq 1$. Additionally, the sum of all diagonal elements is equal to zero since
\begin{align}
    \sum_{\alpha=1}^{N} \omega^{(\alpha-1)} = 0.
\end{align} This completes the proof of Lemma 1.

\section{Proof of Lemma 3}\label{subsec:lemma3}
Given an angle $\phi$, from the definition in \eqref{eq:SINR}, the signal term and the interference-plus-noise term are respectively defined as
\begin{align}
    P_k(\phi) &\eqdef \sqrt{p_{\rm t}}\sv(k)g(\alpha_k\av(\phi),\Fm_k,\yv_k),\\
    I_k(\phi) &\eqdef \sqrt{p_{\rm t}}\sum_{k'\neq k}^{N}\sv(k')v(\alpha_k\av(\phi),\Fm_{k'},\yv_k) \nonumber\\
    &+ ({\bf 1}_N\oslash(\alpha_k\av(\phi))^{*})^{\transp}\Fm_{k}^{*}\zv_k.
\end{align} From the definitions in \eqref{eq:signal} and \eqref{eq:isignal}, it holds that for $\sin\phi \neq \sin\theta_k$,
\begin{align}
    |P_k(\phi)|^2 &< |P_k(\theta_k)|^2\label{eq:powerphi} \\ |I_k(\phi)|^2 &> |I_k(\theta_k)|^2.\label{eq:inequal}
\end{align} First of all, the inequality in \eqref{eq:powerphi} comes from the fact that the combining gain of the desired signal in \eqref{eq:signal} can be represented by
\begin{align}
    |g(\alpha_k\av(\phi),\Fm_k,\yv_k)| &= |({\bf 1}_N\oslash(\alpha_k\av(\phi))^{*})^{\transp}(\alpha_k\av(\theta_k))^{*}|\nonumber\\
    &=|((\av(\phi))^{\herm}\av(\theta_k))^{*}| \stackrel{(a)}{\leq} N,
\end{align} where the equality in (a) holds if and only if $\sin\phi = \sin\theta_k$. On the other hand, the inequality in \eqref{eq:inequal} comes from the fact that the combining gain of the interference signals is given by
\begin{align}
    |v(\alpha_k\av(\phi),\Fm_{k'},\yv_k)| &= |(\av(\phi))^{\transp}\mbox{diag}((\av(\theta_k))^{*})\uv_{[k,k']}^{*}|\nonumber\\
    &\stackrel{(a)}{\geq} 0,
\end{align} where the equality in (a) holds if and only if $\sin\phi = \sin\theta_k$. For $\phi\in[-\pi,\pi)$, therefore, we can get:
\begin{equation}
    \gamma(\alpha_k\av(\phi),\Fm_k,\yv_k)\leq\gamma(\alpha_k\av(\theta_k),\Fm_k,\yv_k),
\end{equation} where the equality holds if and only if $\sin\theta_k = \sin\phi$. This completes the proof of Lemma 3.



\end{document}

%% file: macros.tex
\long\def\comment#1{}

\newfont{\bbb}{msbm10 scaled 700}

\newfont{\bb}{msbm10 scaled 1100}
\newcommand{\CC}{\mbox{\bb C}}

\newcommand{\EE}{\mbox{\bb E}}

\newcommand{\av}{{\bf a}}

\newcommand{\fv}{{\bf f}}

\newcommand{\hv}{{\bf h}}

\newcommand{\pv}{{\bf p}}

\newcommand{\sv}{{\bf s}}

\newcommand{\uv}{{\bf u}}

\newcommand{\vv}{{\bf v}}
\newcommand{\xv}{{\bf x}}
\newcommand{\yv}{{\bf y}}
\newcommand{\zv}{{\bf z}}

\newcommand{\Am}{{\bf A}}
\newcommand{\Bm}{{\bf B}}
\newcommand{\Cm}{{\bf C}}

\newcommand{\Fm}{{\bf F}}

\newcommand{\Id}{{\bf I}}

\newcommand{\Pm}{{\bf P}}

\newcommand{\Um}{{\bf U}}

\newcommand{\Ac}{{\cal A}}

\newcommand{\Cc}{{\cal C}}

\newcommand{\Nc}{{\cal N}}

\newcommand{\Rc}{{\cal R}}

\newcommand{\Wc}{{\cal W}}

\newcommand{\SNR}{{\sf SNR}}

\newcommand{\eqdef}{\stackrel{\Delta}{=}}

\newcommand{\herm}{{\sf H}}

\newcommand{\transp}{{\sf T}}
